\documentclass[11pt,a4paper]{article}
\usepackage{epsfig,amsmath,amssymb,scalefnt,cite}

\long\def\symbolfootnote[#1]#2{\begingroup%
\def\thefootnote{\fnsymbol{footnote}}\footnote[#1]{#2}\endgroup}

\newcommand{\abbrev}{\rm\scalefont{.9}}

\newcommand{\sm}{{\abbrev SM}}
\newcommand{\thdm}{{\abbrev 2HDM}}
\newcommand{\mssm}{{\abbrev MSSM}}
\newcommand{\nmssm}{{\abbrev NMSSM}}
\newcommand{\cp}{{\abbrev CP}}
\newcommand{\nnlo}{{\abbrev NNLO}}
\newcommand{\nnnlo}{{\abbrev NNNLO}}
\newcommand{\nlo}{{\abbrev NLO}}
\newcommand{\lo}{{\abbrev LO}}
\newcommand{\lhc}{{\abbrev LHC}}

\newcommand{\lep}{{\abbrev LEP}}

\newcommand{\qcd}{{\abbrev QCD}}
\newcommand{\sqcd}{{\abbrev SQCD}}
\newcommand{\susy}{{\abbrev SUSY}}

\newcommand{\cms}{{\abbrev cms}}

\newcommand{\sushi}{{\tt SusHi}}

\newcommand{\smallz}{{\scriptscriptstyle Z}} 

\newcommand{\smallr}{{\scriptscriptstyle R}} %
\newcommand{\smallf}{{\scriptscriptstyle F}} %
\newcommand{\mz}{m_\smallz}

\newcommand{\muF}{\mu_\smallf}
\newcommand{\muR}{\mu_\smallr}

\def\mbo{m_{\tilde{b}_1}}
\def\mbt{m_{\tilde{b}_2}}
\def\mto{m_{\tilde{t}_1}}
\def\mtt{m_{\tilde{t}_2}}
\def\ctt{c_{2\theta_t}}
\def\stt{s_{2\theta_t}}
\def\cbt{c_{2\theta_b}}
\def\sbt{s_{2\theta_b}}
\def\mg{m_{\tilde{g}}}

\def\ctw{c_{2\theta_W}}

\def\RS{\mathcal{R}^S}
\def\RP{\mathcal{R}^P}
\def\RG{\mathcal{R}^G}

\newcommand{\eqn}[1]{Eq.\,(\ref{#1})}

\newcommand{\fig}[1]{Fig.\,\ref{#1}}

\newcommand{\sct}[1]{Section~\ref{#1}}
\newcommand{\app}[1]{Appendix~\ref{#1}}

\newcommand{\citere}[1]{Ref.~\cite{#1}}
\newcommand{\citeres}[1]{Refs.~\cite{#1}}

\begin{document}

\begin{titlepage}

 {\flushright{
        \begin{minipage}{2.5cm}
          DESY 15-029
        \end{minipage}        }

}
\renewcommand{\thefootnote}{\fnsymbol{footnote}}
\vskip 2cm
\begin{center}
{\LARGE\bf Neutral Higgs production at proton colliders\\ in the \cp-conserving{\scalefont{.9} NMSSM}}
\vskip 1.0cm
{\Large  Stefan Liebler}
\vspace*{8mm} \\
{\sl
DESY, Notkestra\ss e 85, \\
22607 Hamburg, Germany}
\end{center}
\symbolfootnote[0]{{\tt e-mail address:}} 
\symbolfootnote[0]{{\tt stefan.liebler@desy.de}}

\vskip 0.7cm

\begin{abstract}
We discuss neutral Higgs boson production through gluon fusion
and bottom-quark annihilation in the \cp-conserving $\mathbb{Z}_3$-invariant Next-to-Minimal
Supersymmetric Standard Model (\nmssm{}) at proton colliders. For gluon fusion we
adapt well-known asymptotic expansions in supersymmetric particles
for the inclusion of next-to-leading order contributions of squarks and gluinos
from the Minimal Supersymmetric Standard Model (\mssm{})
and include electroweak corrections involving
light quarks. Together with the resummation of higher-order
sbottom contributions in the bottom-quark Yukawa coupling for
both production processes
we thus present accurate cross section predictions implemented
in a new release of the code \sushi{}. We elaborate on the
new features of an additional SU$(2)_L$ singlet in the
production of \cp{}-even and -odd Higgs bosons with respect to the \mssm{}
and include a short discussion of theoretical uncertainties.
\end{abstract}
\vfill
\end{titlepage}    

\setcounter{footnote}{0}


\section{Introduction}
\label{sec:intro}

After the discovery of a scalar boson at the
Large Hadron Collider (\lhc{}) \cite{Aad:2012tfa,Chatrchyan:2012ufa}
in 2012 an essential task of particle physicists is to reveal the nature of
the Higgs-like state and thus the nature of electroweak
symmetry breaking. Apart from deviations from the Standard Model~(\sm{}) prediction
of the properties of the found Higgs-like state, further work includes the search for
additional less and/or more massive scalar bosons, which
can nicely be accommodated in supersymmetric models.
The Next-to-Minimal Supersymmetric Standard Model (\nmssm{})
extends the Minimal Supersymmetric
Standard Model (\mssm{}) by an SU$(2)_L$ singlet and allows the
dynamical generation of the $\mu$-term through electroweak symmetry breaking~\cite{Ellwanger:2009dp,Maniatis:2009re}. 
The latter singlet-doublet mixing term in the superpotential lifts
the \mssm{} tree-level upper bound of the Higgs mass given by the $Z$-boson mass.
Thus, the \nmssm{} can easily accommodate the \sm{}-like Higgs boson with a mass close to $125$\,GeV.
Whereas for the calculation of the
\nmssm{} Higgs spectrum and branching ratios various
spectrum generators are available and include higher
orders in perturbation theory (see \sct{sec:sushi}),
the calculation of neutral Higgs production
cross sections did not exceed leading order (\lo{})
in quantum chromodynamics involving squarks and gluinos (\sqcd{})~\cite{King:2012tr}
and did not include electroweak corrections - apart from
private implementations in e.g. {\tt HIGLU}~\cite{Spira:1995mt}.

It is therefore timely to present the missing ingredients
and a code for the calculation of accurate neutral Higgs production cross sections in the \nmssm{},
where the five neutral Higgs bosons are predominantly generated through
gluon fusion and bottom-quark annihilation at a proton collider.
For this purpose we extend the code \sushi{}~\cite{Harlander:2012pb}.
For the time being we restrict our implementation
to the real \nmssm{} without additional \cp{} violation,
such that \cp{}-even $H_1,H_2$ and $H_3$ and \cp{}-odd Higgs bosons
$A_1$ and $A_2$ can be distinguished in the Higgs sector.
Most recent efforts related to Higgs physics at the \lhc{} are
summarized in the reports of the \lhc{} Higgs cross section working group
\cite{Dittmaier:2011ti,Dittmaier:2012vm,Heinemeyer:2013tqa}.
The \sm{} Higgs is mainly produced through gluon
fusion, where the Higgs-gluon coupling is mediated through virtual
top- and bottom-quarks~\cite{Georgi:1977gs}. Higher order
\qcd{} corrections at next-to-leading order (\nlo{})
are of large importance~\cite{Djouadi:1991tka,Dawson:1990zj,Spira:1995rr}.
In the effective theory of a heavy top-quark
the inclusive cross section is known to next-to-next-to-leading order (\nnlo{})
in \qcd{}~\cite{Harlander:2002wh,Anastasiou:2002yz,
Ravindran:2003um}, in addition finite top-quark mass effects at \nnlo{} were
calculated~\cite{Marzani:2008az,Harlander:2009mq,Pak:2009dg,
Harlander:2009my,Pak:2011hs}. Beyond \nnlo{} \qcd{} effects are accessible
through resummation \cite{Catani:2003zt,
Moch:2005ky,Idilbi:2005ni,Idilbi:2006dg,Ravindran:2006cg,Ahrens:2008nc} and
electroweak corrections are known
\cite{Actis:2008ug,Aglietti:2004nj,Bonciani:2010ms}.
Meanwhile next-to-\nnlo{} (\nnnlo{}) \qcd{} contributions
were estimated in the so-called threshold expansion
\cite{Ball:2013bra,deFlorian:2014vta,Anastasiou:2014vaa,Anastasiou:2014lda},
but they are not further considered in this publication.

The \sm{} results for Higgs production through
gluon fusion can be adjusted to the \mssm{} and the \nmssm{}
through a proper reweighting of the Higgs couplings
to quarks. However, the gluon fusion process can also be mediated
through their superpartners, the squarks.
With respect to the \mssm{} the only generically new
ingredient, which goes beyond the projection of the physical Higgs bosons
onto the neutral components of the two Higgs doublets,
are couplings of the \nmssm{} singlet to squarks,
since no couplings of the singlet to quarks or gauge bosons are present
in the tree-level Lagrangian.
It is therefore of importance to include squark contributions
to gluon fusion at the highest order possible, even though they
decrease in size with increasing squark masses.
For the pseudoscalars $A_i$ squark
contributions to gluon fusion are only induced at \nlo{},
which motivates to go beyond just \lo{} squark contributions
for all Higgs bosons.
For this purpose we adapt the works
of \citeres{Degrassi:2010eu,Degrassi:2011vq,Degrassi:2012vt}
for the \mssm{} to present \nlo{} \sqcd{} contributions for the \nmssm{},
which are based on an expansion in terms of heavy supersymmetric particles
taking into account terms
up to $\mathcal{O}(m_{\phi}^2/M^2$), $\mathcal{O}(m_t^2/M^2$), $\mathcal{O}(m_b^2/M^2$)
and $\mathcal{O}(m_Z^2/M^2$), where $m_\phi$ denotes the Higgs mass and
$M$ a generic \susy{} mass.
In contrast to the \mssm{} we are at present only working
in this expansion of inverse \susy{} masses and do not include an expansion in the so-called
VHML, the vanishing Higgs mass limit ($m_\phi\rightarrow 0$) for the \sqcd{} contributions,
as implemented in {\tt evalcsusy}~\cite{Harlander:2003bb,Harlander:2004tp,Harlander:2005if}
or discussed in \citere{Degrassi:2008zj}. In the latter limit higher-order
stop-induced contributions up to \nnlo{} level are known
\cite{Harlander:2003kf,Pak:2010cu,Pak:2012xr}
and were partially included in previous discussions of precise
\mssm{} neutral Higgs production cross sections~\cite{Bagnaschi:2014zla}.
Although for a pure \cp{}-odd singlet component \nnlo{} stop-induced contributions
are the first non-vanishing contributions to the gluon fusion cross section,
we leave an inclusion of these to future work.
For completeness, we add that in the \mssm{} a numerical evaluation of \nlo{} squark/quark/gluino
contributions was also reported in
\citere{Anastasiou:2008rm,Muhlleitner:2010nm}, whereas \citeres{Anastasiou:2006hc,Aglietti:2006tp,Muhlleitner:2006wx}
presented analytic results for the pure squark induced \nlo{} contributions.
Electro-weak contributions to the gluon fusion production process
mediated through light quarks \cite{Aglietti:2004nj,Bonciani:2010ms}
can be adjusted from the \sm{} to the \mssm{}~\cite{Harlander:2012pb}
and similarly to the \nmssm{} and are known to capture
the dominant fraction of electroweak contributions for a light \sm{}-like Higgs
with a mass below the top-quark mass, whereas they are generically small
for larger Higgs masses.

For large values of $\tan\beta$, the ratio of the vacuum expectation values
of the neutral components of the two Higgs doublets, the bottom-quark
Yukawa coupling is enhanced, such that the
bottom sector gets more important for gluon fusion, and the associated production
with a pair of bottom-quarks $pp\rightarrow b\bar b\phi$ is significantly enhanced.
\sushi{} includes bottom-quark annihilation $b\bar b\rightarrow \phi$, which in
case of non-tagged final state $b$-quarks is a good theoretical approach, since it
resums logarithms through the $b$-parton distribution
functions. The latter process is known as five-flavor scheme (5FS)
up to \nnlo{} \qcd{} \cite{Maltoni:2003pn,Harlander:2003ai}
and can easily be reweighted from the \sm{} to the \mssm{}/\nmssm{} by effective couplings
\cite{Dittmaier:2006cz,Dawson:2011pe}. In the \nmssm{} the singlet does not couple
to quarks at \lo{}, however taking into account the singlet induced component
into the resummation of higher-order sbottom effects is mandatory, since also the
singlet to sbottom couplings are enhanced by $\tan\beta$.

The new release of \sushi{} thus provides gluon fusion cross sections at \nlo{} \qcd{} taking into
account the third generation quarks and their superpartners, the squarks,
for all the five neutral Higgs bosons of the \nmssm{}.
The squark and squark/quark/gluino contributions are implemented in asymptotic
expansions of heavy \susy{} masses. Electro-weak corrections induced by light quarks
through the couplings of the Higgs bosons to $Z$ and $W^\pm$ bosons can be
added consistently like in the \mssm{}. Similarly, the \nnlo{} top-quark induced
contributions are included. In addition, sbottom contributions can be resummed into
an effective bottom-quark Yukawa coupling, also taking into account the additional
singlet to sbottom couplings. The latter also applies to the calculated bottom-quark annihilation
cross section at \nnlo{} \qcd{}. All features \sushi{} provides for the \mssm{} are
available for the \nmssm{} as well, in particular distributions
with respect to the (pseudo)rapidity and transverse momentum of the Higgs boson under consideration
can be obtained. Left for future work is a link to {\tt MoRe-SusHi}~\cite{Harlander:2014uea}
to allow for the calculation of momentum resummed transverse momentum distributions.

We proceed as follows:
We start with a discussion of the theory background in \sct{sec:setup}, where
we elaborate on the \nmssm{} Higgs sector and the calculation of
the gluon fusion cross section.
Then we present the \nlo{} virtual amplitude for gluon fusion as well as the calculation
of bottom-quark annihilation including the resummation of sbottom-induced
contributions to the bottom-quark Yukawa coupling in the \nmssm{}. Subsequently we comment on
the implementation in \sushi{} in \sct{sec:sushi}, before we investigate the
phenomenological features of the singlet-like Higgs boson in the \cp{}-even
and \cp{}-odd sector with regard to Higgs production in
\sct{sec:pheno}. We also include a short discussion
of theoretical uncertainties. Finally, we conclude and present the
Higgs-squark-squark couplings in Appendix~\ref{app:higgssquark}.

\section{Theory background}
\label{sec:setup}

In this section we discuss the
Higgs sector of the \cp-conserving $\mathbb{Z}_3$-invariant \nmssm{}, before we proceed to the resummation
of $\tan\beta$ enhanced sbottom contributions in the bottom-quark Yukawa coupling.
Subsequently we move to the discussion of the Higgs production cross section
in gluon fusion, where we present the adapted formulas for the \nlo{} \sqcd{}
virtual amplitude, and finally comment on the consequences of the additional
singlet to bottom-quark annihilation.

\subsection{The Higgs sector of the \cp-conserving \nmssm{}}

Our notation of the Higgs sector of the \cp-conserving $\mathbb{Z}_3$-invariant \nmssm{} closely follows
\citere{Ender:2011qh}. For \nmssm{} reviews we refer to \citeres{Ellwanger:2009dp,Maniatis:2009re}.
The superpotential can be written in the form
\begin{align}
 W_{\nmssm}=W_{\mssm}-\epsilon_{ab}\lambda\hat S\hat H_d^a\hat H_u^b + \frac{1}{3}\kappa \hat S^3\quad,
\end{align}
where $W_{\mssm}$ equals the superpotential of the \mssm{} without $\mu$-term.
$\hat S$ denotes the additional SU$(2)_L$ singlet superfield compared to the \mssm{}
with the two SU$(2)_L$ doublet superfields $\hat H_d$ and $\hat H_u$. $\epsilon_{ab}$
contracts the SU$(2)_L$ doublet components.
Since the singlet~$\hat S$ is a neutral field,
it induces one additional \cp{}-even and one additional \cp{}-odd neutral Higgs boson as well
as one additional neutralino compared to the \mssm{}.
The soft-breaking terms include the scalar components $H_d, H_u$ and $S$ of the superfields and are given by
\begin{align}
 \mathcal{L}_{\text{soft}}=\mathcal{L}_{\text{soft},\mssm}
 +(\epsilon_{ab}\lambda A_\lambda S H_d^a H_u^b
 - \frac{1}{3}\kappa A_\kappa S^3 + \text{h.c.}) - m_s^2|S|^2\quad.
\end{align}
The soft-breaking mass $m_s$ can be derived from the minimization conditions of the
tadpole equations (in addition to $m_{H_d}^2$ and $m_{H_u}^2$ like in the \mssm{}),
whereas $A_\lambda$ and $A_\kappa$ are usually
considered input parameters. $A_\lambda$ can be alternatively replaced by
the charged Higgs mass~$m_{H^\pm}$ as input parameter.
The neutral components of the Higgs fields are decomposed according to
\begin{align}
 H_d^0=\frac{1}{\sqrt{2}}(v_d+H_d^R+iH_d^I),\quad
 H_u^0=\frac{1}{\sqrt{2}}(v_u+H_u^R+iH_u^I),\quad
 S=\frac{1}{\sqrt{2}}(v_s+S^R+iS^I),
\end{align}
where $v_d,v_u$ and $v_s$ denote the vacuum expectation values (VEVs) and the fields with
indices $R$ and $I$ are the \cp{}-even and \cp{}-odd fluctuations around them.
An effective $\mu$ term is generated through the VEV of the singlet
\begin{align}
\mu=\frac{1}{\sqrt{2}}\lambda v_s\quad,
\label{eq:effmu}
\end{align}
which will be further used within this article.
We do not present the explicit form of the mass matrices here, but
refer to \citere{Ender:2011qh}. We define the
\cp{}-even gauge eigenstate basis $H^R=(H_d^R,H_u^R,S^R)$ and
the \cp{}-odd one $H^I=(H_d^I,H_u^I,S^I$). Whereas in the
former case the mass eigenstates $H_i$ with $i\in\lbrace 1,2,3\rbrace$
are obtained through one rotation
\begin{align}
H_i=\sum_{j=1}^3\RS_{ij}H^R_j\quad,
\end{align}
we perform a prerotation in the \cp{}-odd sector to obtain the \mssm{} pseudoscalar $A$
and the Goldstone $G$ in the form $H'^I=(G,A,S^I)$. The prerotation $\RG$
is given by the ratio of vacuum expectation values $\tan\beta=v_u/v_d$, such
that the mass eigenstates $A_i$ with $i\in\lbrace 1,2\rbrace$ are obtained by
\begin{align}
A_{i} = \sum_{j=1}^3\RP_{i+1,j}H'^I_j
=\sum_{j,k=1}^3\RP_{i+1,j}\RG_{jk}H^I_k \quad \text{with} \quad 
\RG=\begin{pmatrix}c_\beta&-s_\beta&0\\s_\beta&c_\beta&0\\0&0&1\end{pmatrix}\quad,
\end{align}
where $\RP$ is a $(3\times 3)$-matrix, which however only consists
of a $(2\times 2)$-mixing block, whereas $\RP_{i1}=\RP_{1i}=0$ for $i\neq 1$ and $\RP_{11}=1$.
For Higgs production the Goldstone boson does not need to be considered.
In the following we make use of the notation
``singlet-like Higgs boson'', which refers to the \cp{}-even/odd Higgs boson with the
dominant fraction of the singlet component~$S$ in gauge eigenstates.
For this purpose we define the singlet character $|\mathcal{R}^{S}_{i,3}|^2$ 
for $H_i$ and $|\mathcal{R}^{P}_{i+1,3}|^2$ for $A_i$.
In our discussion of cross sections we denote the Higgs boson by the letter $\phi$,
which can be replaced by any of the physical Higgs bosons $H_i$ or $A_i$.
For picking viable scenarios for phenomenological studies we refer to \citere{Bartl:2009an}
for a recipe to obtain positive eigenvalues for the singlet-like \cp{}-even and -odd Higgs
by varying $A_\kappa$ between a minimal and a maximal
value. 

Whereas the singlet component $S$ does not couple to quarks,
$F$-terms induce a coupling of the singlet-like Higgs to squarks, which
is of relevance for Higgs production.
We present the Higgs-squark-squark couplings to the third generation of
squarks in \app{app:higgssquark}. We point out that the singlet component
mixes with the Higgs doublets proportional to $\lambda$ and also the
couplings of the singlet component to squarks are proportional to $\lambda$.
It is thus possible to mostly decouple the singlet component by lowering
the value of the parameter $\lambda$.
The couplings to quarks can be easily translated from the \mssm{}
by the correct projection on the neutral doublet components $H_d^R$, $H_u^R$
and the pseudoscalar~$A$
and yield relative to the \sm
\begin{align}\nonumber
 g^{H_i}_d&=\RS_{i1}\frac{1}{\cos\beta},\qquad
 g^{H_i}_u=\RS_{i2}\frac{1}{\sin\beta}\\\
 g^{A_i}_d&=\RP_{i+1,2}\tan\beta,\qquad
 g^{A_i}_u=\RP_{i+1,2}\frac{1}{\tan\beta}
\end{align}
with $i\in\lbrace 1,2,3\rbrace$ in the \cp-even and $i\in\lbrace 1,2\rbrace$ in
the \cp-odd Higgs sector. The relative strength $g^{\phi}_f$ enters
the Yukawa couplings in the form $Y^\phi_f=\sqrt{2}m_fg^{\phi}_f/v$ with
the vacuum expectation value $v^2=v_d^2+v_u^2$.

\subsection{Resummation of higher-order sbottom contributions}
\label{sec:resummation}

It is well-known in the \mssm{} that $\tan\beta$ enhanced sbottom corrections to the bottom-quark Yukawa coupling
can be treated in an effective Lagrangian
approach~\cite{Carena:1999py,Carena:2002bb,Guasch:2003cv,Noth:2008tw,Noth:2010jy,Mihaila:2010mp} to be resummed.
For the case of the \nmssm{} just taking into account \sqcd{} corrections the effective Lagrangian
can be written in the form \cite{Baglio:2013iia}
\begin{align}
&\mathcal{L}_{\rm eff} = -Y_b \bar b_R\left[H_d^0+\frac{\lambda\Delta_b}{\mu \tan\beta}S^*H_u^{0*}\right]b_L \qquad\text{with}\\ 
&\Delta_b=\frac{2}{3}\frac{\alpha_s}{\pi}\mg\mu\tan\beta I(\mbo^2,\mbt^2,\mg^2)
\quad\text{and}\quad
I(a,b,c)=\frac{ab\log\frac{a}{b}+bc\log\frac{b}{c}+ca\log\frac{c}{a}}{(a-b)(b-c)(a-c)}\quad.
\end{align}
\citere{Baglio:2013iia} additionally presents the inclusion of \susy{} electroweak corrections proportional
to the soft-breaking parameter $A_t$. The inclusion of electroweak
corrections into $\Delta_b$ does not harm our subsequent discussion of \sqcd{} corrections
and can thus always be included in the bottom-quark Yukawa coupling entering gluon fusion
and bottom-quark annihilation.
Apart from a coupling of the bottom-quarks to the gauge eigenstate $H_u^0$ also an
effective coupling to the singlet $S$ can be induced at loop level, the latter being
proportional to $\lambda v_u$ instead of~$\mu$. The sbottom corrections can be absorbed into effective
Yukawa couplings, which read \cite{Baglio:2013iia}
\begin{align}
\label{eq:resummS}
\tilde{g}_b^{H_i}=\frac{g_b^{H_i}}{1+\Delta_b}\left[1+\Delta_b\left(\frac{\RS_{i2}}{\RS_{i1}\tan\beta}+\frac{\RS_{i3}v\cos\beta}{\RS_{i1}v_s}\right)\right]
\end{align}
for the three \cp{}-even Higgs field $H_i$ and
\begin{align}
\label{eq:resummA}
\tilde{g}_b^{A_i}=\frac{g_b^{A_i}}{1+\Delta_b}\left[1+\Delta_b\left(-\frac{1}{\tan^2\beta}-\frac{\RP_{i+1,3}v}{\RP_{i+1,2}v_s\tan\beta}\right)\right]
\end{align}
for the two \cp{}-odd Higgs fields $A_i$.

\subsection{Gluon fusion cross section}
\label{sec:gluonxs}

After our discussion of the \cp-conserving \nmssm{} Higgs sector and the resummation
of sbottom contributions in the bottom-quark Yukawa coupling, we present
the gluon fusion production cross section for a Higgs boson $\phi$, which
can be written in the form \cite{Harlander:2012pb}
\begin{align}
\sigma(pp\rightarrow \phi+X)
= \sigma_0^\phi \left[1 + C^\phi \frac{\alpha_s}{\pi}\right]\tau_\phi
\frac{d\mathcal{L}^{gg}}{d\tau_\phi }
+\Delta \sigma_{gg}^\phi + \Delta \sigma_{gq}^\phi + \Delta \sigma_{q\overline{q}}^\phi\quad,
\label{eq:gluonxs}
\end{align}
with $\tau_\phi = m_\phi^2/s$ and the hadronic centre-of-mass
energy $s$. The factor $\sigma_0^\phi$ includes the \lo{} partonic cross
section. $C^\phi$ encodes \nlo{} terms 
of singular nature in the limit $\hat s\to m_\phi^2$ with 
the partonic centre-of-mass energy $\hat s$. The gluon-gluon luminosity is given by the integral
\begin{align}
 \frac{d\mathcal{L}^{gg}}{d\tau} = \int_\tau^1 \frac{dx}{x} g(x)g(\tau/x)\quad.
\end{align}
The contributions $\Delta\sigma_{gg}^\phi$,
$\Delta \sigma_{gq}^\phi$, and $\Delta \sigma_{q\overline{q}}^\phi$
are the regular terms in the limit $\hat s\to m_\phi^2$ in the
partonic cross section and arise from $gg$, $gq$ and $q\overline{q}$
scattering, respectively.
The \lo{} contribution $\sigma_0^\phi$ is obtained by the formulas
presented in \citere{Harlander:2012pb} using the \nmssm{} couplings of
the Higgs bosons to quarks and squarks. Similarly the contributions
$\Delta \sigma_{xy}^\phi$ are obtained from the \mssm{} by a proper
replacement of the involved Higgs boson to quark and squark couplings.
The factor $C^\phi$ can be decomposed in the form
\begin{align}
\label{eq:cphi}
 C^\phi=2\text{Re}\left[\frac{\Phi^{(2l)}}{\Phi^{(1l)}_\infty}\right]
+\pi^2+\beta_0\log\left(\frac{\muR^2}{\muF^2}\right)\,,
\end{align}
with $\beta_0=11/2-n_f/3$ and $n_f=5$ as well as the factorization and renormalization
scales, $\muF$ and $\muR$ respectively. $\Phi^{(1l)}_\infty$ 
is the \lo{} (one-loop) virtual amplitude in the limit of large stop and sbottom masses.
$\Phi^{(2l)}$ is the \nlo{} (two-loop) virtual amplitude
and equals the form factors~$\mathcal{H}_i^{(2l)}$ for the \cp{}-even Higgs bosons
and $\mathcal{A}_i^{(2l)}$ for the \cp{}-odd Higgs bosons, which
are presented in the following \sct{sec:formsquark} to account for \nlo{} virtual contributions
from quarks and squarks in an appropriate way. In the \mssm{} limit they correspond to the
form factors of \citeres{Degrassi:2010eu,Degrassi:2011vq,Degrassi:2012vt} except
from a constant factor of $-3/4$.
Contrary to the case of the \mssm{} we do not employ
{\tt evalcsusy} \cite{Harlander:2004tp,Harlander:2005if} to obtain the \nlo{} amplitude
in the limit of heavy top-quark and stop masses for the light Higgs,
but use the expanded form factors presented in the following sections instead.
Accordingly our implementation does not (yet) include approximate \nnlo{} stop contributions
as presented in \citere{Bagnaschi:2014zla} for the \mssm{}.
The \nnlo{} top-quark contributions in the heavy top-quark effective theory making use
of \citeres{Harlander:2002wh,Harlander:2002vv} are included according to
Eq.~(29) of \citere{Harlander:2012pb}.

Lastly we comment on the inclusion of the electroweak corrections to
the gluon fusion production cross section.
Similarly to the \mssm{} the full \sm{} \nlo{} electroweak ({\abbrev EW})
corrections~\cite{Actis:2008ug} can be added to the top-quark induced result only,
assuming complete factorization of {\abbrev EW} and \qcd{} effects~\cite{Anastasiou:2008tj}.
We recommend the latter procedure only for a \sm{}-like Higgs boson.
Contrary the inclusion of electroweak corrections due to light
quarks~\cite{Aglietti:2004nj,Bonciani:2010ms}, where the Higgs boson
couples to either the $Z$ or $W^\pm$ boson, can be adjusted to the \mssm{}
and accordingly the \nmssm{} in an appropriate way~\cite{Bagnaschi:2011tu}.
For this purpose the generalized couplings
\begin{align}
g_V^{H_i} = \RS_{i1}\cos\beta + \RS_{i2}\sin\beta
\label{eq:lqew}
\end{align}
of the $i$-th \cp{}-even \nmssm{} Higgs boson 
to the heavy gauge boson $V\in\lbrace W^\pm,Z\rbrace$ need to be inserted
in the formulas of \citere{Harlander:2012pb}.
The missing projection $\RS_{i3}$ on the singlet component
reflects the fact that the singlet does not couple to gauge bosons. 
The \cp{}-odd Higgs bosons
do not couple to gauge bosons either, such that
electroweak corrections due to light quarks are absent.

\subsection{\nlo{} virtual amplitude for gluon fusion}
\label{sec:formsquark}

Regarding the implementation of two loop contributions to gluon fusion, we
closely follow \citeres{Degrassi:2010eu,Degrassi:2011vq,Degrassi:2012vt}
for the \mssm{}, which can be translated
to the \nmssm{}. Their calculation at \nlo{} is based on an asymptotic
expansion in the masses of the supersymmetric particles.
We can project the form factors onto the ones in gauge eigenstates according to
\begin{align}
\mathcal{H}^{(2l)}_i&=-\frac{3}{4}(\RS_{i1}\mathcal{H}_d^{R,(2l)}
+ \RS_{i2}\mathcal{H}_u^{R,(2l)} + \RS_{i3}\mathcal{S}^{R,(2l)})\\
\mathcal{A}^{(2l)}_{i}&=-\frac{3}{4}(\RP_{i+1,2}\mathcal{H}_A^{I,(2l)} + \RP_{i+1,3}\mathcal{S}^{I,(2l)})\quad.
\end{align}
The individual contributions in gauge eigenstates are presented in
\sct{sec:cpeven} for the \cp{}-even and in \sct{sec:cpodd} for
the \cp{}-odd Higgs bosons. We included the constant factor between \citeres{Degrassi:2010eu,Degrassi:2011vq,Degrassi:2012vt}
and our work in the above equations.

\subsubsection{CP-even Higgs bosons}
\label{sec:cpeven}

In this subsection we present the form factors for the \cp{}-even Higgs bosons
in gauge eigenstates\footnote{In the \cp{}-even sector we adapt the \mssm{} results of \citeres{Degrassi:2010eu,Degrassi:2012vt}
to the \nmssm{} by isolating the terms proportional to the $H_d^R/H_u^R$-squark-squark couplings and replacing
them by the $S^R$-squark-squark couplings for the form factor~$\mathcal{S}$.
Similarly we proceed in the \cp{}-odd sector starting from the form factors of \citere{Degrassi:2011vq}
taking into account the prerotation of the \cp{}-odd Higgs mixing matrix.}
\begin{align}\nonumber
 \mathcal{H}_d^{R,(2l)} &= \frac{1}{\sin\beta}\left[ -m_t\mu s_{2\theta_t}F_t^{2l} + \mz^2s_{2\beta}D_t^{2l}\right]
 +\frac{1}{\cos\beta} \left[ m_b A_bs_{2\theta_b}F_b^{2l} + 2m_b^2G_b^{2l} + 2\mz^2c_\beta^2D_b^{2l}\right]\\\nonumber
 \mathcal{H}_u^{R,(2l)} &= \frac{1}{\cos\beta}\left[ -m_b\mu s_{2\theta_b}F_b^{2l} - \mz^2s_{2\beta}D_b^{2l}\right]
 +\frac{1}{\sin\beta} \left[ m_t A_ts_{2\theta_t}F_t^{2l} + 2m_t^2G_t^{2l} - 2\mz^2s_\beta^2D_t^{2l}\right] \\
 \mathcal{S}^{R,(2l)} &= \frac{1}{\sin\beta} \left[ - \frac{1}{\sqrt{2}}m_t \lambda v_d s_{2\theta_t} F_t^{2l}\right]
 +\frac{1}{\cos\beta} \left[ - \frac{1}{\sqrt{2}}m_b\lambda v_u s_{2\theta_b}F_b^{2l}\right]\quad,
\end{align}
which includes the effective $\mu$ parameter defined in \eqn{eq:effmu}. All functions in $\mathcal{H}_d^{R,(2l)}$,
$\mathcal{H}_u^{R,(2l)}$ and $\mathcal{S}^{R,(2l)}$ can be directly taken over from \citeres{Degrassi:2010eu,Degrassi:2012vt}, keeping in
mind the different convention in the sign of the $\mu$~parameter.
For on-shell (OS) parameters (see \citeres{Degrassi:2010eu,Degrassi:2012vt}) and thus
for our implementation
the contribution $F_t^{2l}$ is shifted according to Section 3.3 of \citere{Degrassi:2012vt}
and $F_b^{2l}$ according to \citere{Degrassi:2010eu}. The shift also applies to $F_t^{2l}$ 
and  $F_b^{2l}$ entering the singlet contribution $\mathcal{S}^R$, since the differences
in the prefactors being $\mu$, $\lambda v_d$ or $\lambda v_u$ are not renormalized
when taking into account \sqcd{} contributions
and therefore do not contribute to the described OS shifts.

It remains to discuss the inclusion of resummed sbottom contributions
into the bottom-quark Yukawa coupling within the virtual corrections to gluon fusion,
where care has to be taken to avoid a double-counting of \nlo{} \sqcd{} contributions.
The naive resummation $\tilde{g}_b=g_b/(1+\Delta_b)$ is incorporated in the same way
as in case of the \mssm{} \cite{Degrassi:2010eu,Bagnaschi:2014zla}. The resummation
as presented in \sct{sec:resummation} instead needs the subtraction of the $\tan\beta$ enhanced
contributions to $G_b^{2l}$ multiplied with the corresponding coupling correction, in detail
for the three \cp{}-even Higgs bosons $H_i$
\begin{align}
\label{eq:shiftS}
2m_b^2G_b^{2l}\rightarrow 2m_b^2G_b^{2l}-\frac{C_F}{2}\mathcal{G}_{1/2}^{1l}(\tau_b)\mu\tan\beta\left(-\mg I(\mbo^2,\mbt^2,\mg^2)\right)K'_i
\end{align}
with the factor $K'_i$ being
\begin{align}
K'_i=\frac{1}{1+\Delta_b}\left[1-\left(\frac{\RS_{i2}}{\RS_{i1}\tan\beta}+\frac{\RS_{i3}v\cos\beta}{\RS_{i1}v_s}\right)\right],\qquad i\in\lbrace 1,2,3\rbrace\quad.
\end{align}
All occurrences of the bottom-quark Yukawa coupling in the two-loop amplitude
are multiplied with the factor $K_i=\tilde{g}_b^{H_i}/g_b^{H_i}$ using \eqn{eq:resummS}, such
that the shift reported in \eqn{eq:shiftS} avoids double-counting of the purely \sqcd{} induced contributions
at the two-loop level.
Employing the expansion in heavy \susy{} masses the \nlo{} virtual contributions to neutral \cp{}-even Higgs
production in the \nmssm{} are now fully presented.

\subsubsection{CP-odd Higgs bosons}
\label{sec:cpodd}

We now turn to the case of the two \cp{}-odd Higgs bosons, where we present the form
factor in the basis $H'^I$ after a prerotation from gauge eigenstates\footnotemark[1].
At \lo{} only diagrams involving quarks coupling to the pseudoscalar $A$ exist, such that
the form factor at \lo{} only consists of the part $\mathcal{H}_A^{I,(1l)}$, whereas $\mathcal{S}^{I,(1l)}$ equals zero.
At \nlo{} however couplings of $A$ and $S^I$ to
squarks $\tilde{q}_i\tilde{q}_j$ for $i\neq j$ induce contributions to Higgs production.
The two-loop form factor presented in \citere{Degrassi:2011vq} for the
\mssm{} can therefore be translated to
\begin{align}\nonumber
&\mathcal{H}_A^{I,(2l)}=\left[\cot\beta(\mathcal{K}^{2l}_{tg}+\mathcal{K}^{2l}_{t\tilde{t}\tilde{g}})
+\tan\beta(\mathcal{K}^{2l}_{bg}+\mathcal{K}^{2l}_{b\tilde{b}\tilde{g}})\right]\\
&\mathcal{S}^{I,(2l)}=\left[\cot\beta\mathcal{K}^{S,2l}_{t\tilde{t}\tilde{g}}+\tan\beta\mathcal{K}^{S,2l}_{b\tilde{b}\tilde{g}}\right]\quad.
\end{align}
Whereas the individual contributions to $\mathcal{H}_A^{I,(2l)}$ can be taken from \citere{Degrassi:2011vq},
we present the contributions to $\mathcal{S}^{I,(2l)}$ separately:
\begin{align}\nonumber
\mathcal{K}^{S,2l}_{t\tilde{t}\tilde{g}}=&\frac{C_F}{2}\mathcal{K}^{1l}(\tau_t)\frac{\mg}{m_t}\frac{m_t\frac{1}{\sqrt{2}}\lambda v}{\mto^2-\mtt^2}
\left(\frac{x^t_1}{1-x^t_1}\ln x^t_1-\frac{x^t_2}{1-x^t_2}\ln x^t_2\right)\\
&-\frac{m_t}{\mg}\stt\mathcal{R'}^t_1
+\frac{2m_t^2 \frac{1}{\sqrt{2}}\lambda v}{\mg(\mto^2-\mtt^2)}\mathcal{R}_2
-\frac{1}{2}\mathcal{K}^{1l}(\tau_t)\frac{m_{A_i}^2}{\mto^2-\mtt^2}\mathcal{R'}^t_4\\
\mathcal{K}^{S,2l}_{b\tilde{b}\tilde{g}}=&\frac{C_F}{2}\mathcal{K}^{1l}(\tau_b)\frac{\mg}{m_b}\frac{m_b\frac{1}{\sqrt{2}}\lambda v}{\mbo^2-\mbt^2}
\left(\frac{x^b_1}{1-x^b_1}\ln x^b_1-\frac{x^b_2}{1-x^b_2}\ln x^b_2\right)-\frac{m_b}{\mg}\sbt\mathcal{R'}^b_1
\end{align}
with $x^t_i=m_{\tilde{t}_i}^2/\mg^2$ and $x^b_i=m_{\tilde{b}_i}^2/\mg^2$. The functions $\mathcal{K}^{1l}$
and $\mathcal{R}_2$ can be found in \citere{Degrassi:2011vq}.
The functions $\mathcal{R'}^t_1$ and $\mathcal{R'}^t_4$ are given by:
\begin{align}\nonumber
\mathcal{R'}^t_1=&\frac{C_F}{(x^t_1-x^t_2)^2}\frac{\frac{1}{\sqrt{2}}\lambda v}{\mg}
\left(1+\frac{1}{2}\mathcal{K}^{1l}(\tau_t)\right)\left[\frac{x^{t2}_1(1-2x^t_2)}{2(1-x^t_1)(1-x^t_2)}
+\frac{x^t_1(x^{t2}_1-2x^t_2+x^t_1x^t_2)\ln x^t_1}{2(1-x^t_1)^2}\right]\\
&-(x^t_1\leftrightarrow x^t_2)\\
\mathcal{R'}^t_4=&\frac{C_F}{(x^t_1-x^t_2)^2}\frac{\frac{1}{\sqrt{2}}\lambda v}{\mg}
\left[\frac{x^{t2}_1(1-2x^t_2)}{2(1-x^t_1)(1-x^t_2)}+\frac{x^t_1(x^{t2}_1-2x^t_2+x^t_1x^t_2)\ln x^t_1}{2(1-x^t_1)^2}\right]-(x^t_1\leftrightarrow x^t_2)
\end{align}
In the bottom sector the relevant function yields:
\begin{align}
\mathcal{R}'^b_1=\frac{C_F}{(x^b_1-x^b_2)^2}\frac{\frac{1}{\sqrt{2}}\lambda v}{\mg}
\left[\frac{x^{b2}_1(1-2x^b_2)}{2(1-x^b_1)(1-x^b_2)}+\frac{x^b_1(x^{b2}_1-2x^b_2+x^b_1x^b_2)\ln x^b_1}{2(1-x^b_1)^2}\right]-(x^b_1\leftrightarrow x^b_2)
\end{align}
The shifts of individual contributions in case of OS parameters can be taken
over from the \mssm{} case.
The inclusion of resummed sbottom contributions to the bottom-quark Yukawa coupling needs
the following shift in the two-loop form factor for the \cp{}-odd Higgs bosons~$A_i$
\begin{align}
\mathcal{K}^{2l}_{b\tilde{b}\tilde{g}} \rightarrow 
\mathcal{K}^{2l}_{b\tilde{b}\tilde{g}}
 - \frac{C_F}{2}\mathcal{K}^{1l}(\tau_b)\mu\tan\beta\left(-\mg I(\mbo^2,\mbt^2,\mg^2)\right)K'_i
\end{align}
using
\begin{align}
K'_i=\frac{1}{1+\Delta_b}\left[1+\left(\frac{1}{\tan^2\beta}+\frac{\RP_{i+1,3}v}{\RP_{i+1,2}v_s\tan\beta}\right)\right],\qquad i\in\lbrace 1,2\rbrace\quad.
\end{align}
Again we point out that our sign convention with respect to $\mu$ is opposite to \citere{Degrassi:2011vq}
and all occurrences of the bottom-quark Yukawa coupling in the two loop amplitude are
multiplied with the factor $K_i=\tilde{g}_b^{A_i}/g_b^{A_i}$ using \eqn{eq:resummA}.

\subsection{Bottom-quark annihilation cross section in the 5FS}
\label{sec:bottomquarkan}

The generalization of the calculation of bottom-quark annihilation cross sections
in the five-flavor scheme (5FS) from the \mssm{} to the \nmssm{} case is straightforward
by using the appropriate couplings of Higgs bosons to bottom-quarks.
For this purpose the resummation of sbottom contributions
as described in \sct{sec:resummation} is taken
into account. For the specific case of the singlet-like Higgs boson we point out
that in case the coupling to the bottom-quark vanishes (due to cancellations
in the mixing with the Higgs doublets) a priori the coupling to sbottom
squarks can still be present. This is not taken into account by the resummation
procedure.

\section{Implementation in \sushi{}}
\label{sec:sushi}

In the current implementation of neutral Higgs production in the real \nmssm{}
within the code \sushi{} the Higgs mixing matrices as well as the Higgs
masses have to be provided as input in {\tt \susy{} Les Houches Accord (SLHA})
form~\cite{Skands:2003cj,Allanach:2008qq} and can be obtained by
spectrum generators for the \nmssm{}. Common codes are
{\tt NMSSMTools}~\cite{Ellwanger:2004xm,Ellwanger:2005dv,Belanger:2005kh,Ellwanger:2006rn},
{\tt NMSSMCALC}~\cite{Djouadi:1997yw,Graf:2012hh,Ender:2011qh,Baglio:2013iia,Muhlleitner:2014vsa},
{\tt SOFTSUSY}~\cite{Allanach:2001kg,Allanach:2013kza},
{\tt SPheno+Sarah}~\cite{Staub:2013tta,Porod:2003um,Porod:2011nf} and
{\tt FlexibleSUSY+Sarah}~\cite{Staub:2013tta,Athron:2014yba}.

Special attention needs to be paid to the renormalization of the stop and sbottom sector,
which in the ideal form should be identical in the calculation of
Higgs masses and mixing and the calculation of Higgs production cross
sections. For the time being, \sushi{} either relies on the internal calculation
of on-shell stop and sbottom sectors as described in the manual \cite{Harlander:2012pb}
or on the specification of the on-shell masses $m_{\tilde{q}_1}$
and $m_{\tilde{q}_2}$ and mixing angles $\theta_{\tilde{q}}$
in the input file. For both cases input files can be found
in the folder {\tt /example} within the \sushi{} tarball.
The user is asked to check the meaning of output parameters
of spectrum generators, i.e. the chosen renormalization scheme.
If the user specifies the on-shell squark masses and
mixing angles together with the on-shell soft-breaking parameters $A_t$
and $A_b$ by hand, she/he should make sure that in the stop sector
$A_t$ as well as the on-shell top-quark mass $m_t$, the on-shell stop masses $\mto$
and $\mtt$ and the mixing angle $\theta_{\tilde{t}}$ fit the formula
\begin{align}
 \sin(2\theta_{\tilde{t}}) = \frac{2m_t(A_t-\mu/\tan\beta)}{\mto^2-\mtt^2}\quad.
\end{align}
In the sbottom sector on-shell and tree-level masses
on the other hand differ by a shift in the $(1,1)$-element,
see $\Delta M_L^2$ in \citere{Harlander:2012pb}.
Moreover we employ the scheme which works with a dependent bottom-quark mass $m_b$,
whereas $A_b$ is defined to be on-shell, see e.g.
\citeres{Brignole:2002bz,Heinemeyer:2004xw,Heinemeyer:2010mm}.
To allow for maximal flexibility
the specification of on-shell squark masses and mixing angles is now also possible
in case of the \mssm{}. The {\tt Block RENORMSBOT} is
not of relevance in such input files, since $m_b$ is chosen
as dependent parameter, whereas the squark mixing angle $\theta_b$ and
the soft-breaking parameter $A_b$
are understood as renormalized on-shell.

Two options for the pseudoscalar Higgs mixing matrix
are accepted as input by \sushi{}, namely the full Higgs mixing
matrix, which corresponds to the multiplication $\RP\RG$ in the
above notation, but instead also the rotation matrix $\RP$ can
be used as input. Following {\tt SLHA2}~\cite{Allanach:2008qq}
the full matrix $(\RP\RG)_{ij}$ is provided
in {\tt Block NMAMIX} and asks for entries $ij$ with $i\in\lbrace 2,3\rbrace$
and $j\in\lbrace 1,2,3\rbrace$. The matrix $\RP_{ij}$ can be specified
in {\tt Block NMAMIXR}, which only asks for entries $ij$ with
$\lbrace i,j\rbrace \in \lbrace 2,3\rbrace$. We point out that
in contrast to other codes the Goldstone boson remains
the first mass eigenstate, such that {\tt Block NMAMIXR}
does not ask for entries with $i=1$ or $j=1$.
The elements of the \cp{}-even Higgs boson mixing matrix are specified in
{\tt Block NMHMIX}~\cite{Allanach:2008qq}.
The Higgs masses
need to be given in {\tt Block MASS} using entries
$25,35$ and $45$ for the \cp{}-even Higgs bosons and
$36$, $46$ for the \cp{}-odd Higgs bosons.

The block {\tt Block EXTPAR} still contains the gluino mass
as well as the soft-breaking parameters for the
third generation squark sector. Entry $23$ for the $\mu$ parameter is however
replaced by entry $65$, where the effective value of $\mu$ needs
to be specified. Moreover entry $61$ asks for the choice of $\lambda$.
\sushi{} extracts the VEV $v_s$ from $\mu$ and $\lambda$.
Since the Higgs sectors including their mixing are provided,
there is no need to provide the parameters $\kappa$, $A_\kappa$, $A_\lambda$
(or $m_{H^\pm}$) in the \sushi{} input, since they do not enter the
couplings relevant for Higgs production. 
The {\tt Block SUSHI} entry $2$ specifies the Higgs boson,
for which cross sections are requested. The \cp{}-even Higgs
bosons are numbered $11,12$ and $13$, the \cp{}-odd Higgs
bosons $21$ and $22$. Similarly the options $11,12$ and $21$ also work
in the 2-Higgs-Doublet Model (\thdm{}) and the \mssm{} and $11$ and $21$ in the \sm{}.
A \cp{}-odd Higgs boson $21$ in the \sm{} is obtained from
the \thdm{} case with $\tan\beta=1$.
We note that \sushi{} is still compatible with input files
with $0$ (light Higgs), $1$ (pseudoscalar) and $2$ (heavy Higgs)
as options for entry $2$. Output files however stick to the
new convention.

For the time being we emphasize that \sushi{} is not strictly suitable
for very low values of Higgs masses $m_\phi<20$\,GeV, where 
quark threshold effects start to become relevant and also
electroweak corrections are not implemented.
This statement mostly applies to studies of a very light \cp{}-odd Higgs boson, which is
poorly constrained by \lep{} experiments in contrast
to a light \cp-even Higgs boson~\cite{Schael:2006cr}.

\vspace{-1mm}
\section{Phenomenological study}
\label{sec:pheno}

In this section we elaborate on the phenomenological consequences of the additional
SU$(2)_L$ singlet in the \nmssm{} with respect to the \mssm{} for neutral Higgs production.
Neglecting the squark induced contributions to gluon fusion, the only consequence of the additional
singlet component is another admixture of the three \cp{}-even/two \cp{}-odd Higgs bosons. However,
no generically new contributions to Higgs boson production arise.
This differs when taking into account squark induced contributions to gluon fusion 
due to the additional singlet to squark couplings. In particular for the
\cp{}-odd Higgs bosons squark contributions are only induced
at the two-loop level due to the non-diagonal structure of the \cp{}-odd Higgs bosons to squark couplings.
Subsequently we work with two scenarios, start with their definition, present
the Higgs boson masses and admixtures and then discuss the behavior of
cross sections, including the squark and electroweak corrections to the
gluon fusion cross section. Our studies are performed for a proton-proton collider
with a centre-of-mass (\cms{}) energy of $\sqrt{s}=13$\,TeV, as planned
for the second run of the \lhc{}.
Lastly we add a short discussion of renormalization
and factorization scale uncertainties as well as PDF$+\alpha_s$ uncertainties
for one of the two scenarios.

\subsection{Scenarios $S_1$ and $S_2$}

To present the most relevant features of the \nmssm{} for what concerns
neutral Higgs production we pick two scenarios. The first scenario $S_1$ is in the vicinity
of the natural \nmssm{}~\cite{King:2012tr} with a rather large value of $\lambda=0.62$.
Other input parameters are $M_1=150$\,GeV, $M_2=340$\,GeV, $M_3=1.5$\,TeV,
$\tan\beta=2$, $A_\kappa=-20$\,GeV
and $\mu=200$\,GeV. $A_\lambda$ is determined from the charged Higgs
mass $m_{H^\pm}=400$\,GeV.
The size of $\lambda$ ensures a large mixing of the singlet component
with the $H_d^0$ and $H_u^0$ doublets.
All soft-breaking masses are set to $1.5$\,TeV except
for the soft-breaking masses of the third generation squark sector, which
are fixed to $750$\,GeV. The soft-breaking couplings are set to $A=1.8$\,TeV.
The on-shell stop masses are then given by $\mto=544.7$\,GeV
and $\mtt=941.2$\,GeV, whereas the sbottom masses are $\mbo=749.4$\,GeV
and $\mbt=757.4$\,GeV.
We vary $\kappa$ between $0.15$ and $0.80$
and thus vary the mass of the singlet-like Higgs component in particular
in the \cp{}-even Higgs sector. We note that for illustrative reasons
the perturbativity limit approximately given by $\sqrt{\lambda^2+\kappa^2}<0.7$ is not
always fulfilled in our study.
We work out the characteristics
for the singlet-like component in the following discussion.
The relevant input for {\tt SusHi} is obtained
with {\tt NMSSMCALC 1.03}, which incorporates the leading two-loop
corrections $\mathcal{O}(\alpha_s\alpha_t)$ to the Higgs boson masses
calculated in the gaugeless limit with vanishing external momentum~\cite{Muhlleitner:2014vsa}.
We request {\tt NMSSMCALC} to work with an on-shell renormalized stop sector
and add local modifications to the {\tt NMSSMCALC} input routines to read in on-shell
parameters rather than $\overline{\text{DR}}$ renormalized parameters\footnote{We thank Kathrin Walz for
instructions on how to modify the {\tt NMSSMCALC} input routines.}.
These modifications guarantee identical on-shell stop masses
in {\tt NMSSMCALC} and \sushi{}. The renormalization of the sbottom
sector on the other hand is performed \sushi{}-internally.

We also choose a second scenario $S_2$, in which we vary $\lambda$ to decouple
the singlet-like Higgs from the Higgs doublets.
The detailed choice of parameters is $M_1=150$\,GeV, $M_2=300$\,GeV,
$M_3=1.5$\,TeV, $\tan\beta=10$, $A=-2.0$\,TeV, $\kappa=0.2$, $A_\kappa=-30$\,GeV, $\mu=130$\,GeV
and $m_{H^\pm}=350$\,GeV. In this scenario we set the soft-breaking masses
to $1.0$\,TeV.
The on-shell stop and sbottom masses are given by $\mto=824.1$\,GeV,
$\mtt=1173.4$\,GeV, $\mbo=998.0$\,GeV and $\mbt=1008.4$\,GeV.
We vary $\lambda$ between $0.04$ and $0.25$. For small values of $\lambda$
$H_1$ corresponds to the \sm{}-like Higgs boson with mass $m_{H_1} \sim 121$\,GeV.
The lower bound at $\lambda=0.04$ is to avoid tiny cross sections for a heavy singlet-like Higgs boson
and to keep its mass below the \susy{} masses thresholds to justify the \nlo{} \sqcd{} expansion
employed for the gluon fusion cross section calculation.

Both our scenarios come along with rather light third generation squark masses
at the low TeV scale.
Contrary to the Higgs mass calculations the squark contributions completely decouple
from Higgs production for heavy \susy{} spectra. Our scenarios are
chosen to flash the phenomenology of an additional singlet-like Higgs boson
and thus do not always include a \sm{}-like Higgs boson with mass $\sim 125$\,GeV and are
partially under tension from \lep{} searches~\cite{Schael:2006cr} (for low
\cp{}-even Higgs masses below $110$\,GeV) or
\lhc{} searches
\cite{Chatrchyan:2012am,Chatrchyan:2013qga,Aad:2013hla,Aad:2014dya,Chatrchyan:2014tja,Aad:2014ioa,
Khachatryan:2014wca,Aad:2014vgg,Khachatryan:2014jya,Aad:2015wra,CMS:2014onr,CMS:2014yra,Aad:2014iia,atlas:conf2014071,CMS:2014cdp}.


We add for both scenarios the relevant \sm{} input, which includes the $\overline{\text{MS}}$ renormalized 
bottom-quark mass $m_b(m_b)=4.20$\,GeV, which is translated into a bottom-quark pole mass
of $m_b=4.92$\,GeV. In \sushi{} we choose the renormalization scheme, where the bottom-quark pole mass
enters all occurrences of heavy bottom-quark masses in the loops and the bottom-quark Yukawa coupling
for the gluon fusion cross section. Bottom-quark annihilation is based on the running 
$\overline{\text{MS}}$ renormalized bottom-quark Yukawa coupling.
As pointed out in \citere{Bagnaschi:2014zla}
the gluon densities are hardly dependent on the bottom-quark pole mass fit value of the PDF fitting groups,
emphasizing that there is no need to adjust the bottom-quark pole mass to the PDF fit value
for the calculation of the gluon fusion cross section.
The top-quark pole mass equals $m_t=173.3$\,GeV. The strong coupling
constant~$\alpha_s(m_Z)$ is set to $0.1172$ for the calculation of running masses, and is obtained
from the corresponding PDF set for the cross section calculation. We choose
{\tt MSTW2008}~\cite{Martin:2009iq} at the appropriate order in perturbation theory.
Our central scale choices for gluon fusion are $m_\phi/2$ for both renormalization
and factorization scale, $\muR^0$ and $\muF^0$ respectively, and
$\muR^0=m_\phi$ and $\muF^0=m_\phi/4$ for bottom-quark annihilation.

\subsection{Higgs boson masses and singlet admixtures}

\begin{figure}[htp]
\begin{center}
\begin{tabular}{cc}
\includegraphics[width=0.47\textwidth]{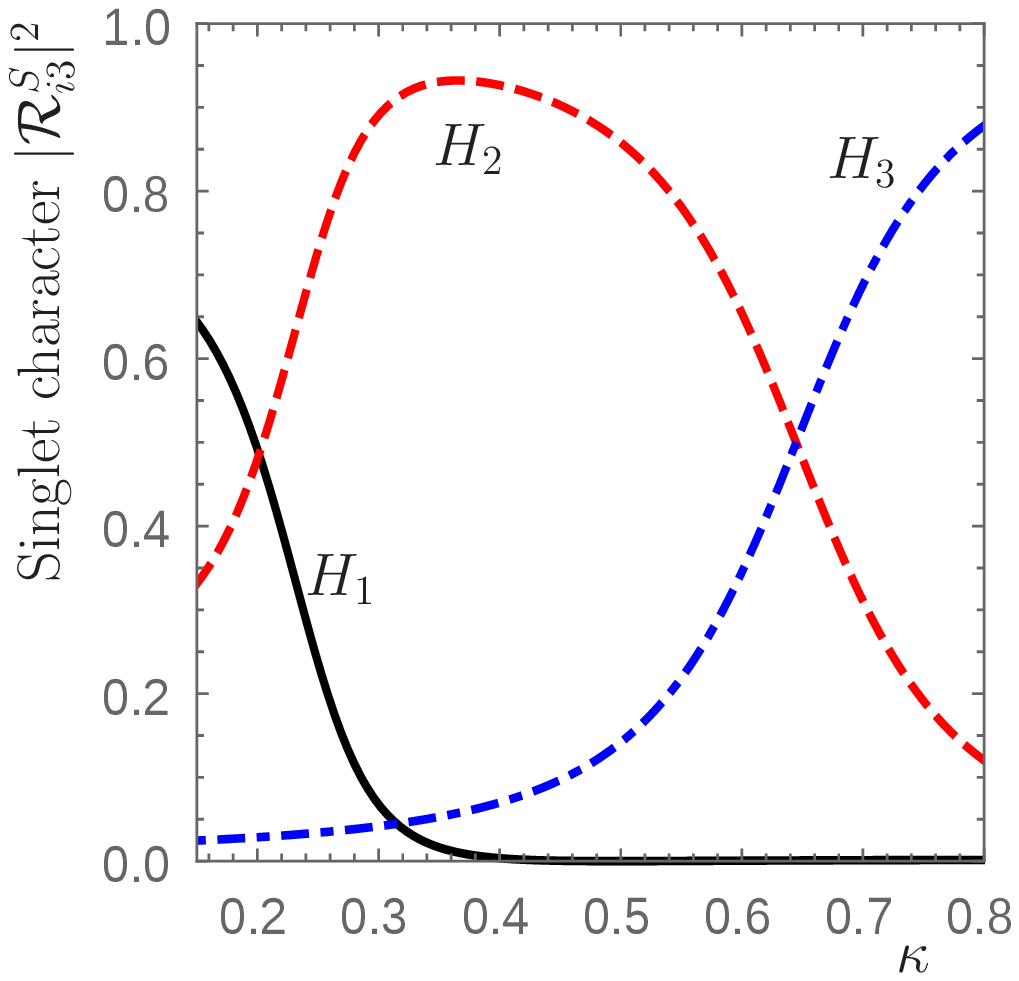} & 
\includegraphics[width=0.47\textwidth]{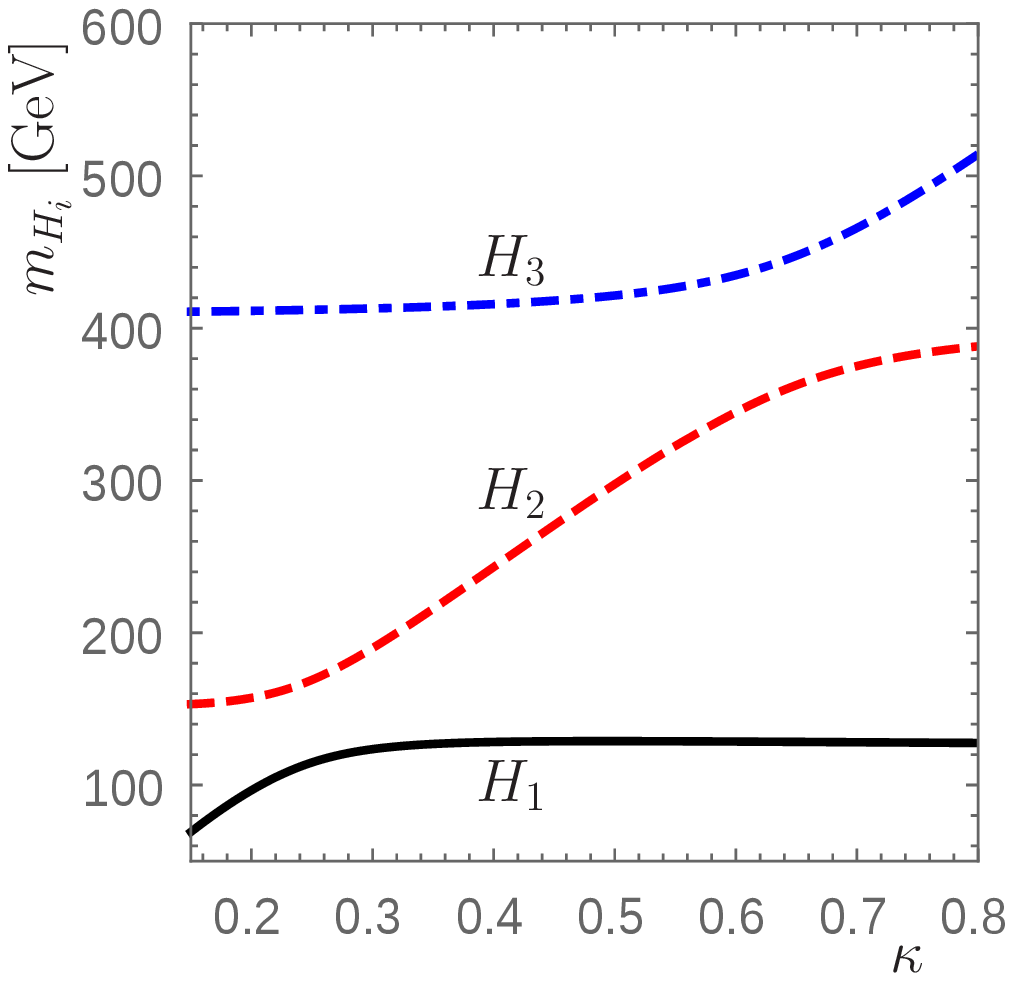}  \\[-0.4cm]
 (a)  & (b) \\
\includegraphics[width=0.47\textwidth]{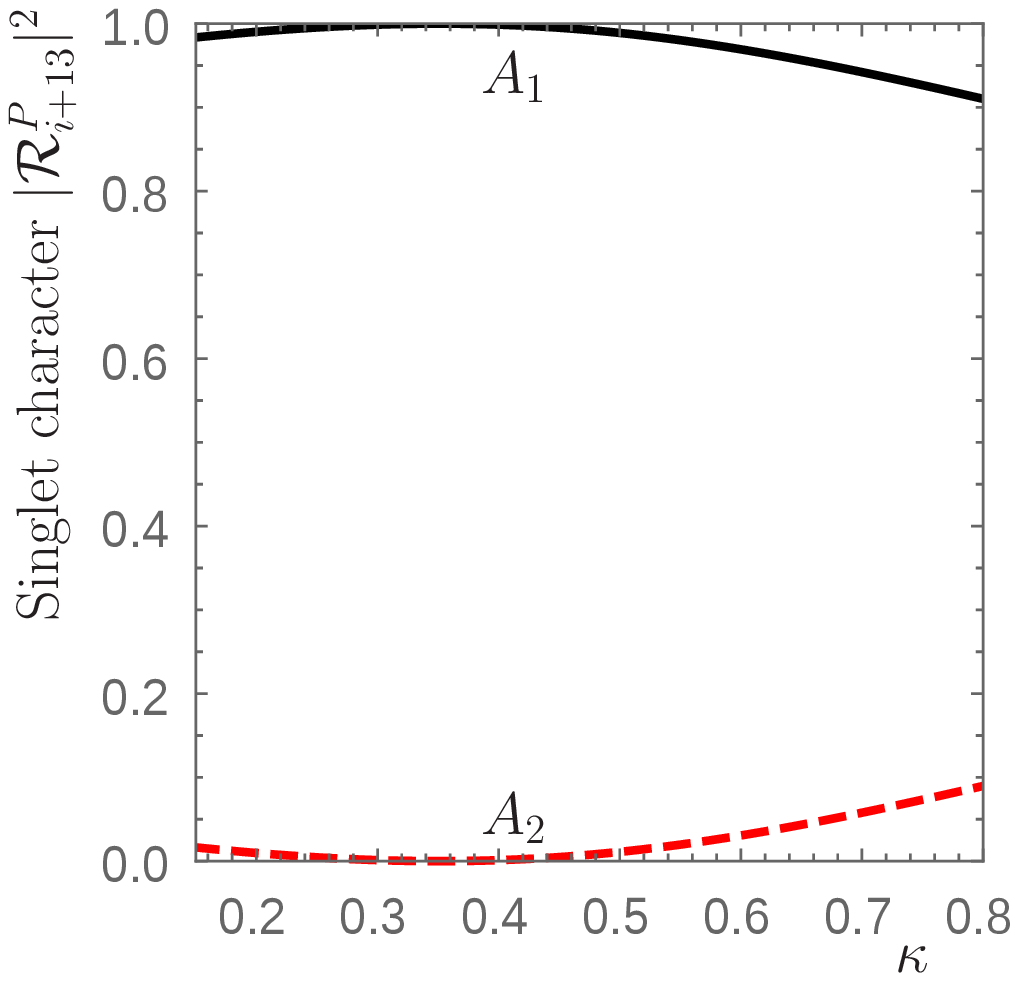} & 
\includegraphics[width=0.47\textwidth]{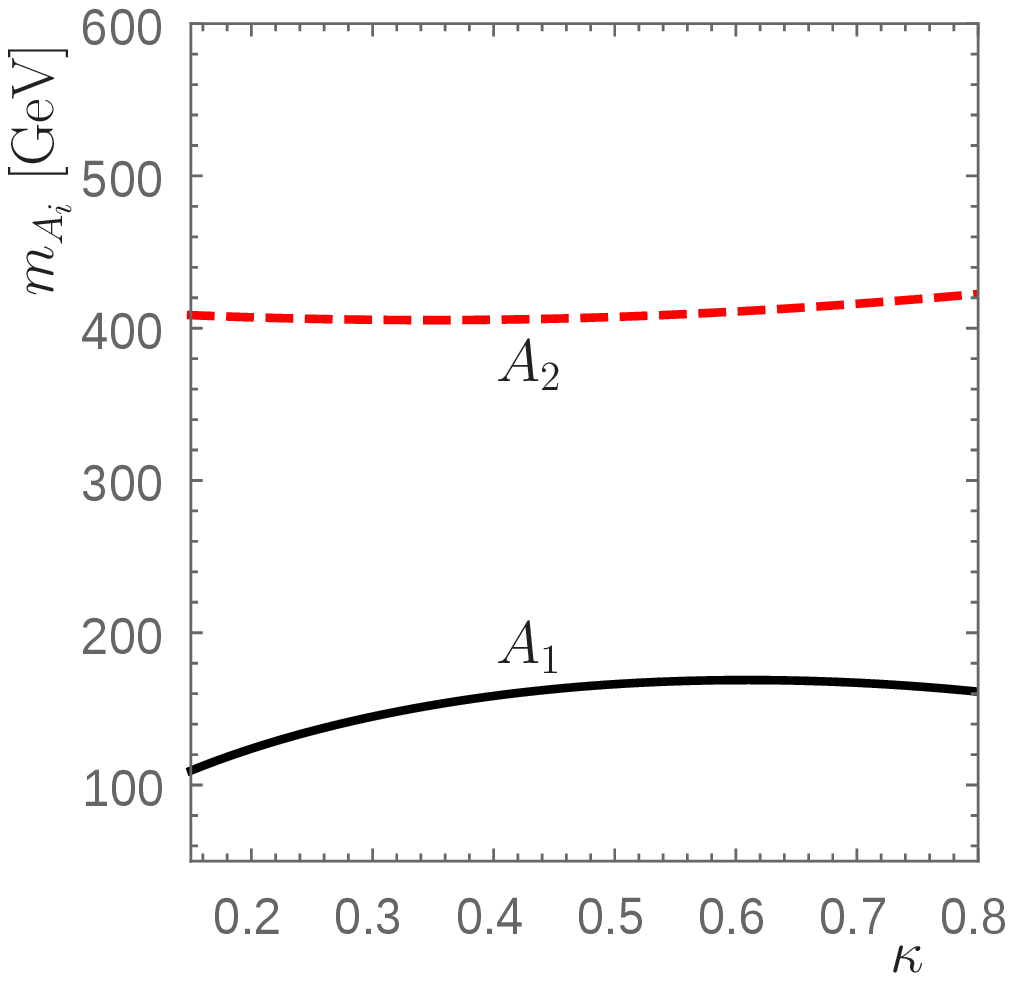}  \\[-0.4cm]
 (c)  & (d) 
\end{tabular}
\end{center}
\vspace{-0.6cm}
\caption{(a) Singlet character and (b) masses of the three \cp{}-even Higgs bosons
$H_1$ (black), $H_2$ (red, dashed), $H_3$ (blue, dotdashed) as a function of $\kappa$
for scenario~$S_1$ obtained from {\tt NMSSMCALC 1.03};
(c) Singlet character and (d) masses of the two \cp{}-odd Higgs bosons $A_1$ (black), $A_2$ (red, dashed)
as a function of $\kappa$
for scenario~$S_1$ obtained from {\tt NMSSMCALC 1.03}.}
\label{fig:CPdefine} 
\end{figure}

Subsequently we start with a discussion of the singlet admixture and the
masses of the three \cp{}-even and the two \cp{}-odd Higgs bosons, which
we obtain through a link to {\tt NMSSMCALC~1.03} as explained beforehand.
For scenario $S_1$ the singlet component as a function of $\kappa$ for the three \cp{}-even
Higgs bosons is shown in \fig{fig:CPdefine}~(a). Clearly, for low values
of $\kappa$ the lightest Higgs $H_1$ is mainly singlet-like, whereas
with increasing $\kappa$ the dominant singlet fraction moves from $H_1$ to $H_2$
and for large values of $\kappa$ to $H_3$. The sum of 
all singlet components yields $\sum_i |\RS_{i3}|^2=1$.
The masses of the \cp{}-even Higgs bosons can be found in \fig{fig:CPdefine}~(b).
With increasing $\kappa$ the mass term of the singlet component in gauge eigenstates
is increasing proportional to $\kappa v_s$, such that the singlet-like
Higgs boson can be identified with the Higgs boson linearly increasing in mass.
Close to $\kappa\sim 0.35$ $H_2$ shows the most
dominant singlet fraction, which will later be visible in the gluon
fusion cross section.
Scenario $S_1$ includes for $\kappa>0.3$ a \sm{}-like Higgs boson $H_1$
with a mass of $m_{H_1}\sim 125$\,GeV. 
For very small values of $\kappa$ the decay $H_2\rightarrow H_1H_1$ opens and 
leaves a characteristic signature for the \sm{}-like Higgs boson $H_2$.
Note that a light singlet-like Higgs boson lifts the mass
of the \sm{}-like \cp{}-even Higgs through singlet-doublet mixing,
which for our example equals $m_{H_2}\sim 153$\,GeV for $\kappa=0.1$.
The region of small $\kappa$ and a light \cp{}-even singlet-like Higgs boson~$H_1$
is largely constrained by the \lep{} experiments~\cite{Schael:2006cr}.

\fig{fig:CPdefine}~(c) and (d) show the behavior of the singlet admixture and the masses
for the two \cp{}-odd Higgs bosons in scenario $S_1$ as a function of $\kappa$.
We point again to the region in the vicinity of $\kappa\sim 0.35$,
where the light \cp{}-odd Higgs boson~$A_1$
is a pure singlet-like \cp{}-odd Higgs boson contrary to the \cp{}-even Higgs boson $H_2$,
for which $H_d^R$ and $H_u^R$ components remain.
The coupling of $A_1$ to quarks vanishes, but the
coupling to squarks is still present due to the relatively large value
of $\lambda=0.62$, which will be apparent when calculating the
gluon fusion cross section.

\begin{figure}[htp]
\begin{center}
\begin{tabular}{cc}
\includegraphics[width=0.47\textwidth]{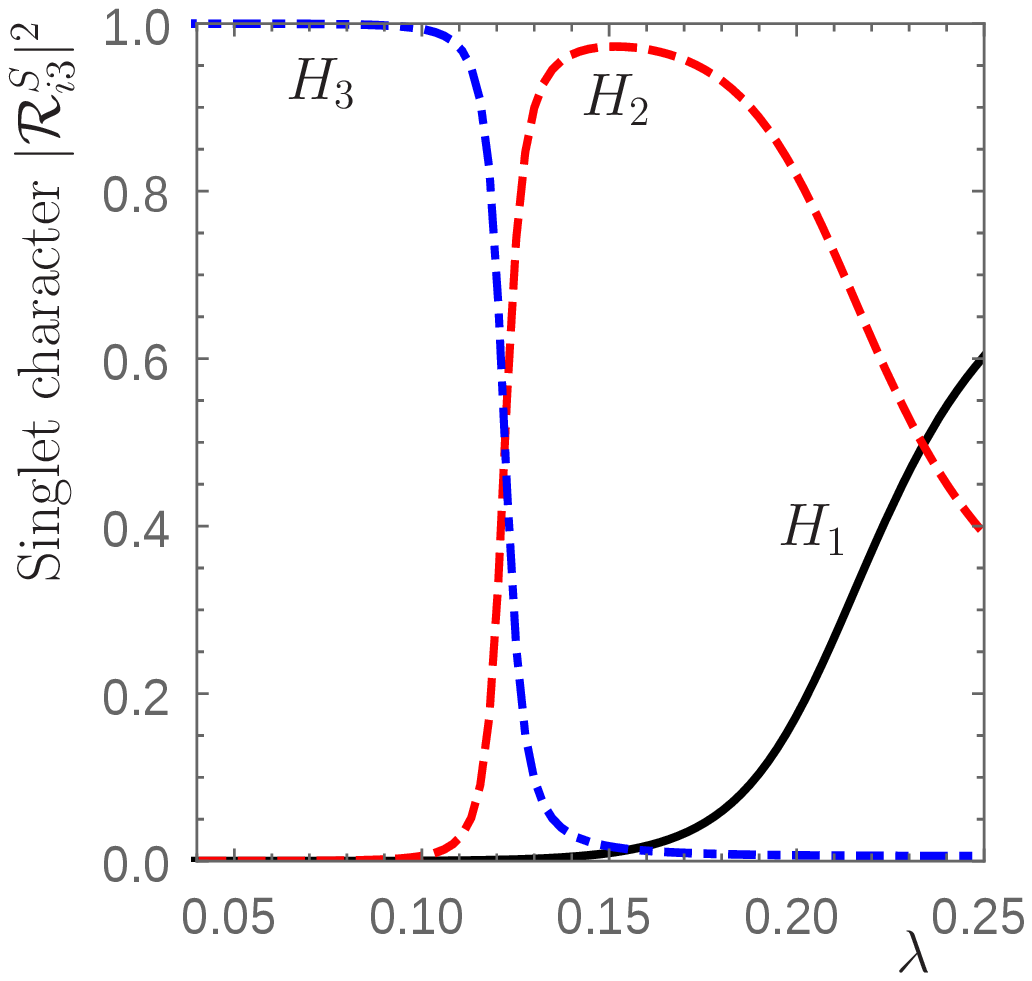} & 
\includegraphics[width=0.47\textwidth]{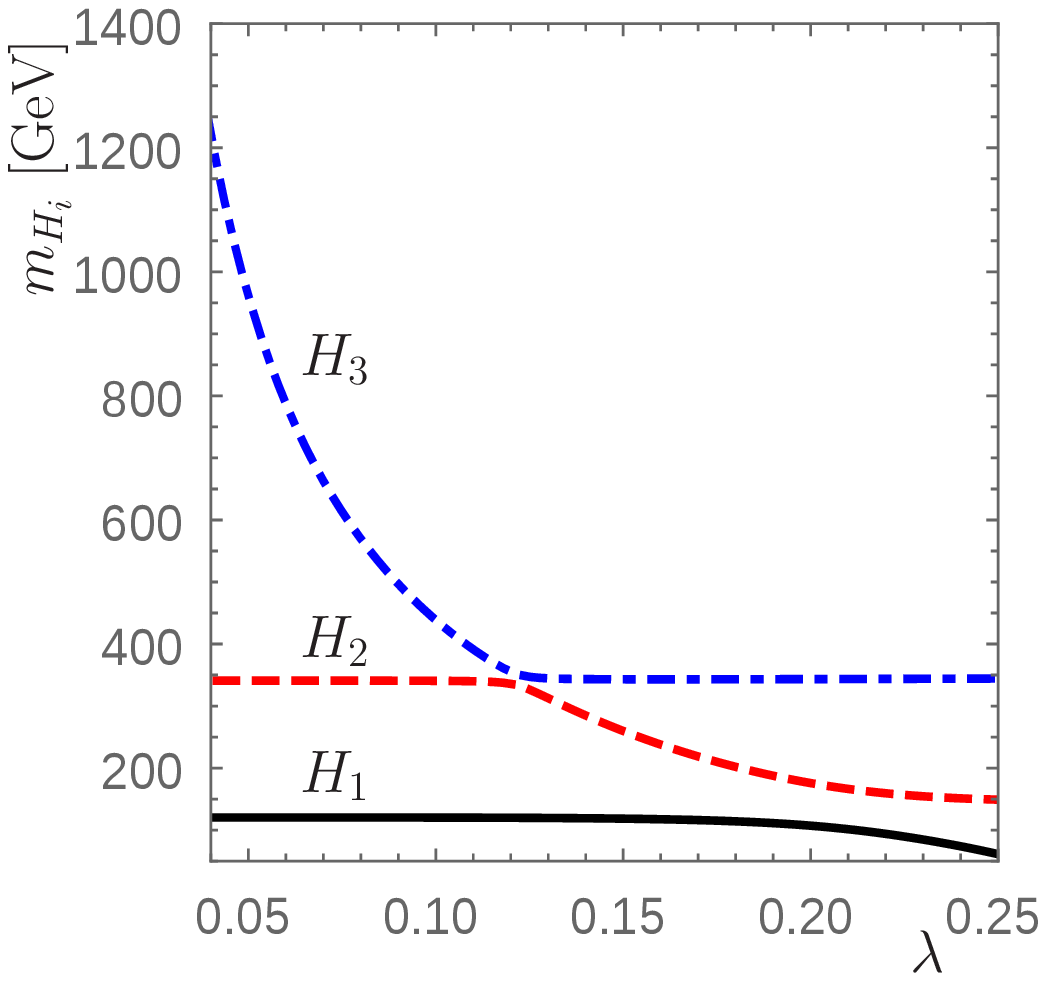}  \\[-0.4cm]
 (a) & (b)\\
\includegraphics[width=0.47\textwidth]{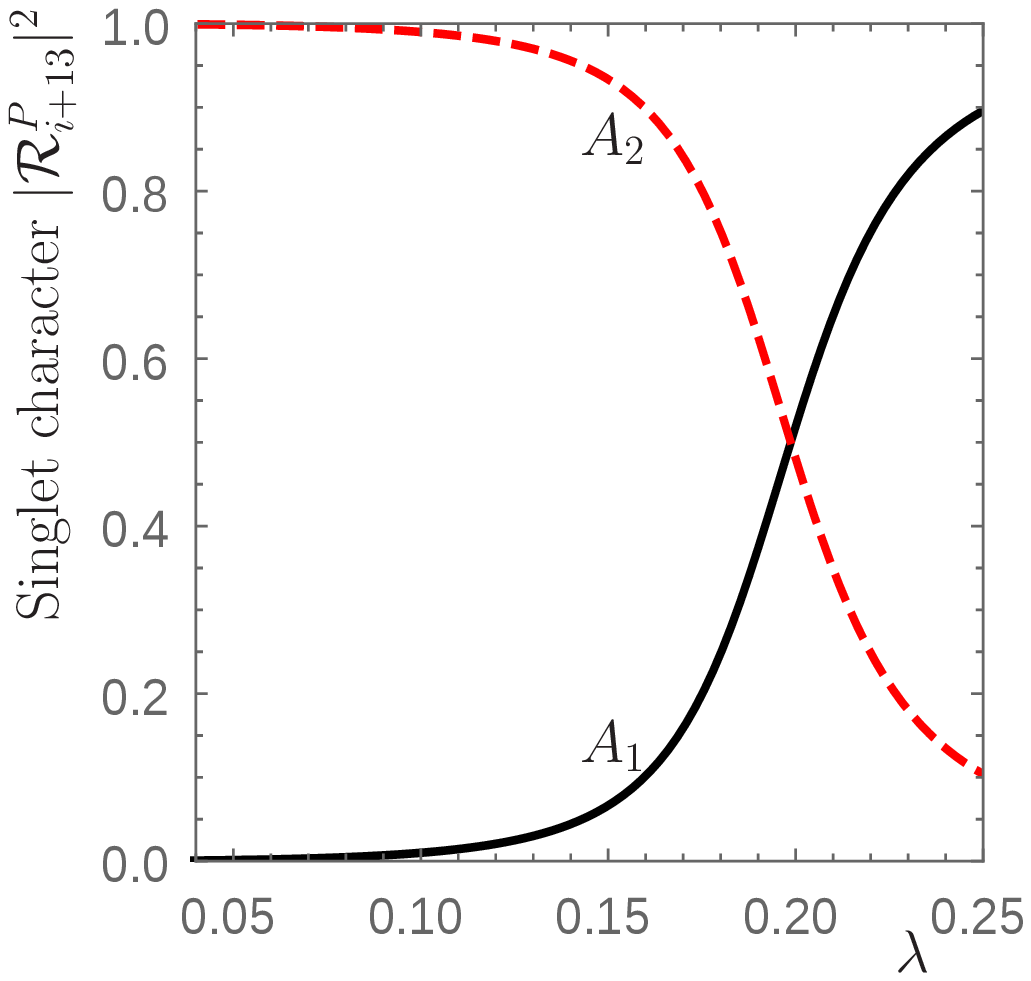} & 
\includegraphics[width=0.47\textwidth]{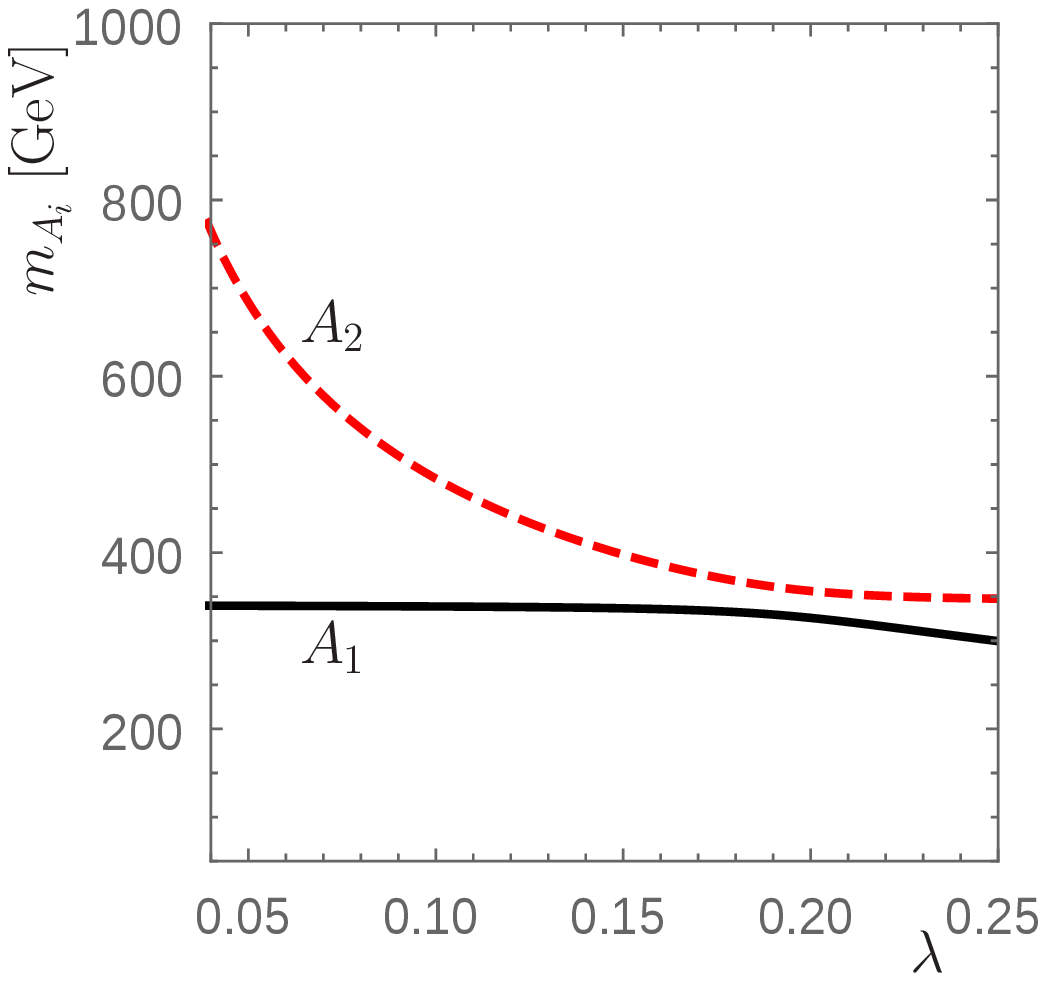}  \\[-0.4cm]
 (c) & (d)
\end{tabular}
\end{center}
\vspace{-0.6cm}
\caption{(a) Singlet character and (b) masses of the three \cp{}-even Higgs bosons
$H_1$ (black), $H_2$ (red, dashed), $H_3$ (blue, dotdashed) as a function of $\lambda$
for scenario~$S_2$ obtained from {\tt NMSSMCALC 1.03};
(c) Singlet character and (d) masses of the two \cp{}-odd Higgs bosons $A_1$ (black), $A_2$ (red, dashed)
as a function of $\lambda$
for scenario~$S_2$ obtained from {\tt NMSSMCALC 1.03}.}
\label{fig:CPodefine} 
\end{figure}

For scenario $S_2$ \fig{fig:CPodefine} shows correspondingly the singlet character
and masses for the \cp{}-even and \cp{}-odd Higgs bosons. Due to the fixed
value of $\mu=\tfrac{1}{\sqrt{2}}\lambda v_s$ the singlet-like Higgs
boson increases in mass (proportional to $\kappa v_s$) with decreasing $\lambda$
and thus for small $\lambda$
$H_3$ as well as $A_2$ clearly decouple from the other Higgs bosons. We will
later use this setup to show the decoupling behavior of the cross sections.
Below $\lambda<0.05$ both $H_3$ and $A_2$ have a singlet character, which
exceeds $|\mathcal{R}^{S/P}_{33}|^2>0.999$.

\subsection{Scenario $S_1$: Inclusive cross sections for $\sqrt{s}=13$\,TeV}

\begin{figure}[htp]
\begin{center}
\begin{tabular}{cc}
\includegraphics[width=0.47\textwidth]{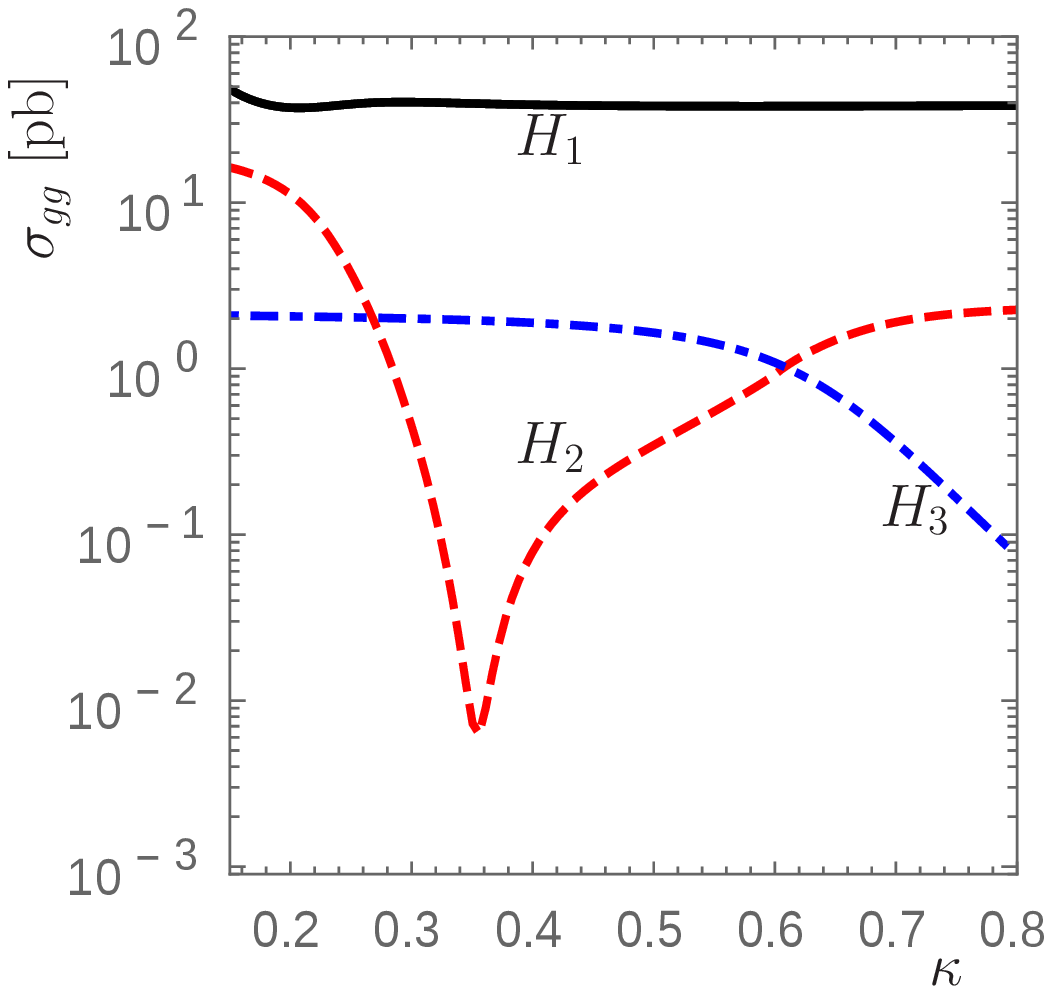} & 
\includegraphics[width=0.47\textwidth]{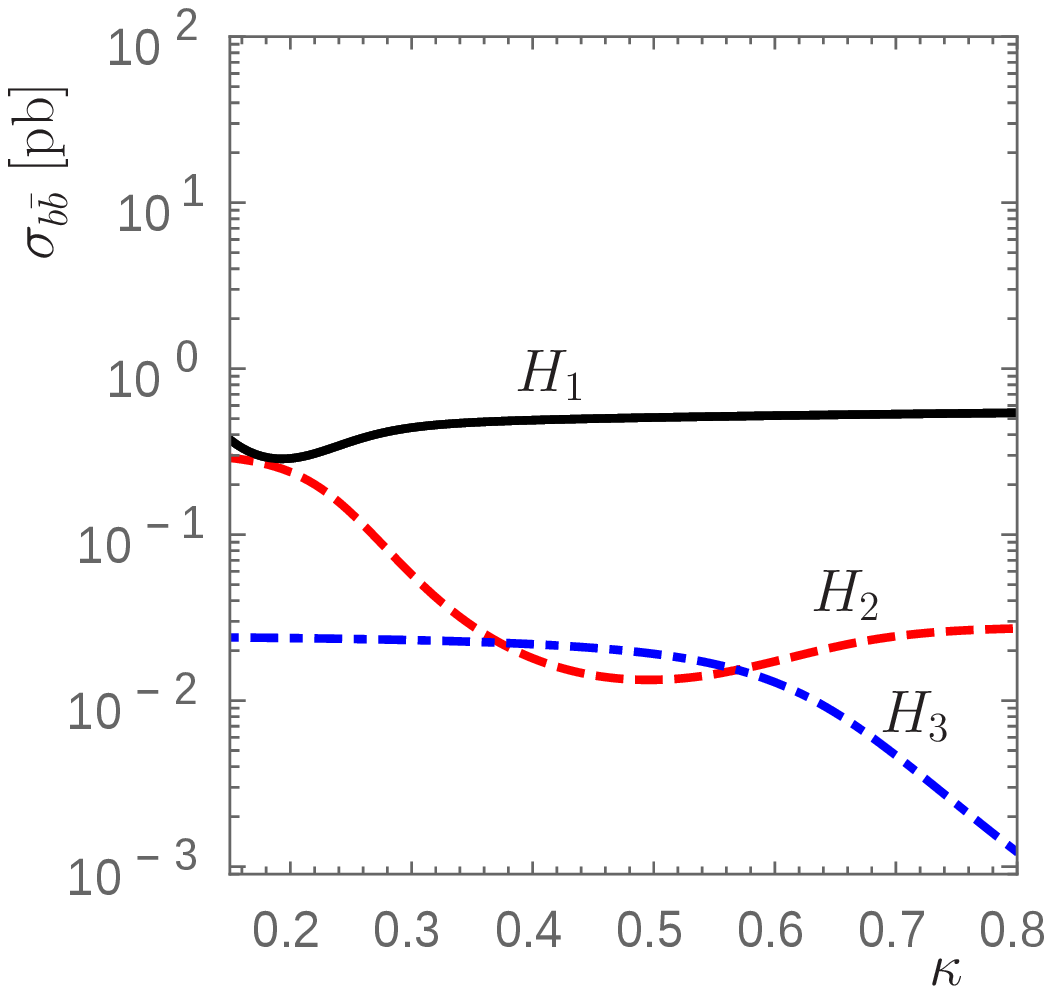}  \\[-0.4cm]
 (a) & (b)  \\
\includegraphics[width=0.47\textwidth]{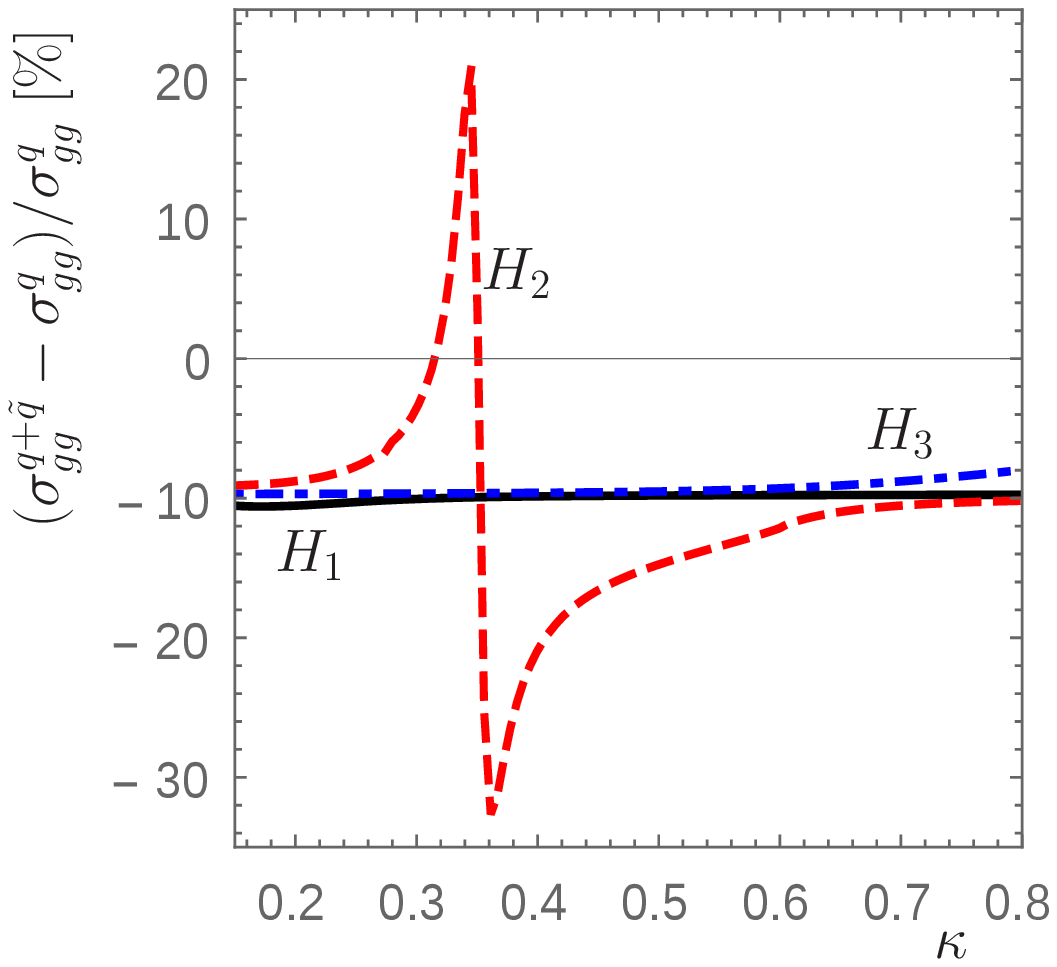} & 
\includegraphics[width=0.47\textwidth]{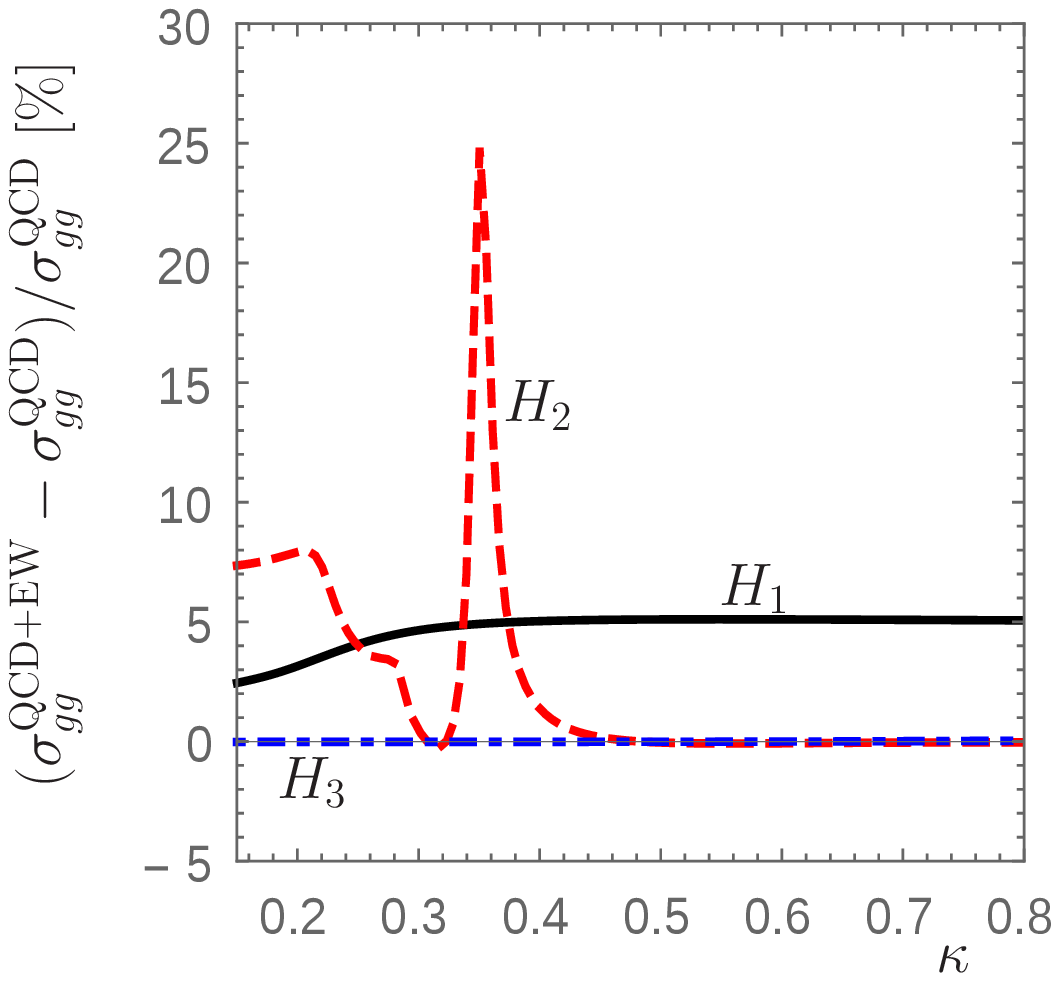}  \\[-0.4cm]
 (c)  & (d) 
\end{tabular}
\end{center}
\vspace{-0.6cm}
\caption{(a) Gluon fusion and (b) bottom-quark annihilation in pb at $\sqrt{s}=13$\,TeV
as well as (c) squark and (d) electroweak contributions to gluon fusion
for the three \cp{}-even Higgs bosons $H_1$ (black), $H_2$ (red, dashed), $H_3$ (blue, dotdashed)
as a function of $\kappa$ for scenario~$S_1$.}
\label{fig:CPeggh} 
\end{figure}

In this subsection we investigate
the gluon fusion $\sigma_{gg}$ and bottom-quark annihilation $\sigma_{b\bar b}$ cross sections
for scenario $S_1$ for $\sqrt{s}=13$\,TeV for a proton-proton collider.
The subsequent statements are however hardly dependent on the \cms{} energy
and thus hold for the $7/8$\,TeV \lhc{} runs as well
as for more energetic runs.
\fig{fig:CPeggh} shows the cross sections for the three \cp{}-even Higgs bosons.
Naturally the cross sections are strongly dependent on the Higgs mass,
which are in turn a function of $\kappa$. Thus, the cross section
for the second \cp{}-even Higgs bosons $H_2$ tends to decrease
with increasing $\kappa$. Crucial is the singlet admixture
of the Higgs boson under consideration.
The larger the singlet component $|\RS_{i3}|^2$, the smaller the coupling
to quarks becomes and thus the more sensitive is the cross section
to squark and electroweak contributions. For $H_2$ we observe a cancellation of
quark contributions through the admixtures with the SU$(2)_L$ doublets
around $\kappa\sim 0.35$, where in turn due to the generally small
cross section squark
but also electroweak corrections to the gluon fusion cross section
are of large relevance, see \fig{fig:CPeggh}~(c) and (d). 
For small values of $\kappa$ the decay $H_2\rightarrow H_1H_1$
opens in addition to the large gluon fusion cross section for the singlet-like
\cp{}-even Higgs boson $H_1$. The region is therefore constrained
by \lep{} experiments~\cite{Schael:2006cr}.
Much smoother is the behavior for the bottom-quark annihilation cross section,
where the direct coupling to bottom-quarks is related to the
non-singlet character of the Higgs under consideration.
In an interval around $\kappa\sim 0.35$ bottom-quark annihilation
even exceeds the gluon fusion cross section for $H_2$
despite the small value of $\tan\beta=2$.

We show the effect of squark and electroweak contributions to gluon fusion
for the three \cp{}-even Higgs bosons in \fig{fig:CPeggh}~(c) and (d).
$\sigma_{gg}^{q+\tilde{q}}$ in \fig{fig:CPeggh}~(c) includes stop- and
sbottom-quark induced contributions at \nlo{} \sqcd{} on top of the
quark induced contributions without electroweak
contributions and compares to the pure quark induced cross section $\sigma_{gg}^q$
without electroweak contributions.
All cross sections include \nlo{} \qcd{} quark contributions and the
\nnlo{} \qcd{} top-quark induced contributions in the heavy top-quark effective theory.
\fig{fig:CPeggh}~(d) accordingly
shows the effect of electroweak contributions induced by light quarks following
\eqn{eq:lqew} in combination with \citere{Bagnaschi:2014zla}
in comparison to the quark and squark induced cross section
$\sigma_{gg}^{\qcd{}}=\sigma_{gg}^{q+\tilde{q}}$. Note that in all our
figures $\sigma_{gg}$ corresponds to $\sigma_{gg}^{\qcd+\rm{EW}}$.
As expected for $H_2$ the region with small quark contributions induced by the admixture with
the $H_d^R$ and $H_u^R$ components is in particular sensitive to squark corrections.
For the other Higgs bosons the squarks corrections in this scenario are incidentally
all of the order of $\mathcal{O}(-10$\%) and mostly independent of $\kappa$. We note that
the squark corrections are mainly induced by stop contributions,
whereas sbottom-induced contributions only account for a small fraction.
Interestingly, the squark contributions show an interference-like structure
with a maximum and minimum around $\kappa\sim 0.35$, whereas the
relative electroweak corrections are always positive.
This can be understood from a sign change in the real part of the quark induced
\lo{} and \nlo{} amplitude for $H_2$ at $\kappa\sim 0.35$, which is of relevance for
the squark contributions, whereas the imaginary part, more relevant for the electroweak
contributions, does not change its sign.
The size of the electroweak corrections for $H_2$ follows from a suppression
of the couplings of the second lightest Higgs $H_2$ to quarks in contrast
to the couplings to gauge bosons. Obtaining a pure singlet-like Higgs boson
in the \cp{}-even Higgs sector, which neither couples to quarks and gauge bosons,
rarely happens due to the mixing between both $S^R$ and
$H_d^R$ as well as $S^R$ and $H_u^R$ for large values of $\lambda$.
For the \sm{}-like Higgs boson with a mass below the top-quark mass the
electroweak corrections by light quarks are typically of the order of $\mathcal{O}(+5$\%) and cover
most of the \sm{}-electroweak correction factor. On the other hand, for Higgs masses above
the thresholds $m_\phi \gg 2m_W$ or
$2m_Z$ the electroweak corrections by light quarks are small.
The structure visible for $H_2$ in \fig{fig:CPeggh}~(c) for $\kappa<0.3$ is induced
by the thresholds $2m_W$ and $2m_Z$, which the Higgs mass $m_{H_2}$
crosses between $\kappa=0.1$ and $0.3$.
We leave the distortion of distributions, in particular transverse
momentum distributions, for such a scenario to future studies.

\begin{figure}[htp]
\begin{center}
\begin{tabular}{ccc}
\includegraphics[width=0.31\textwidth]{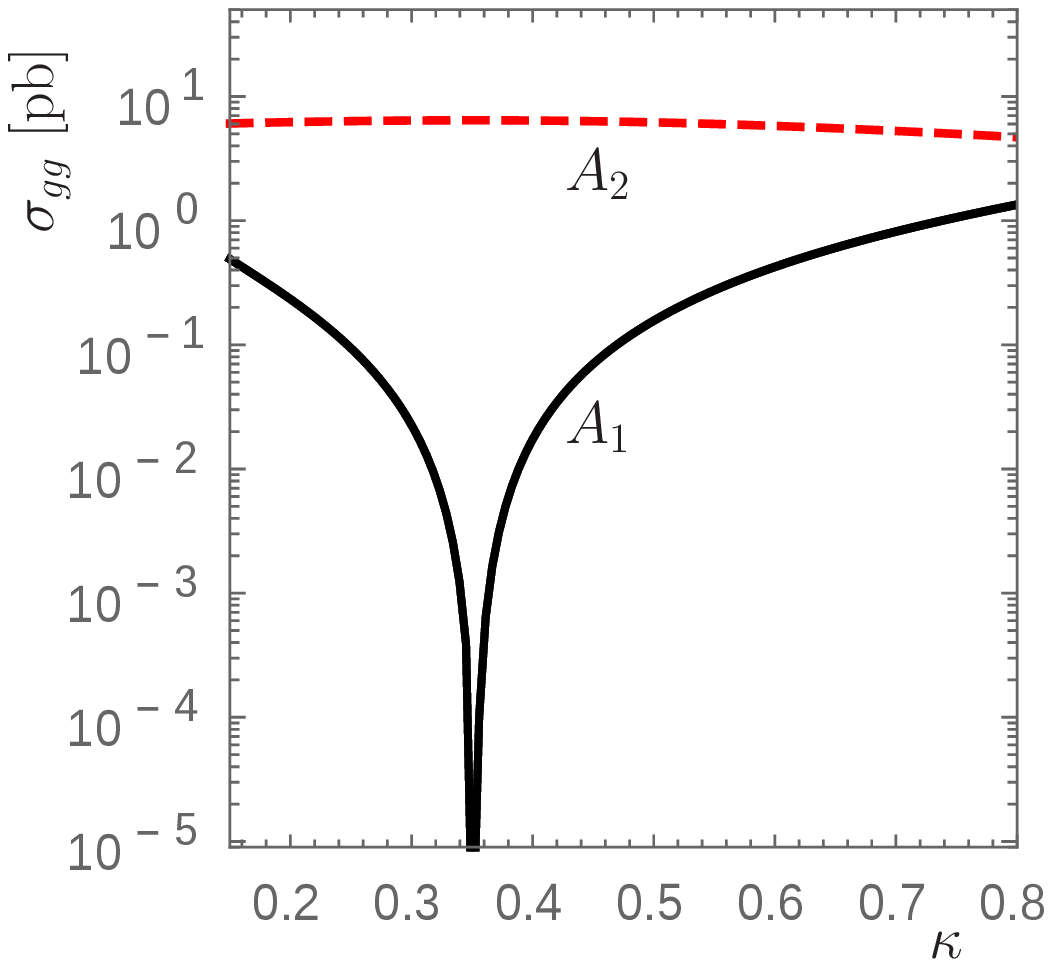} & 
\includegraphics[width=0.31\textwidth]{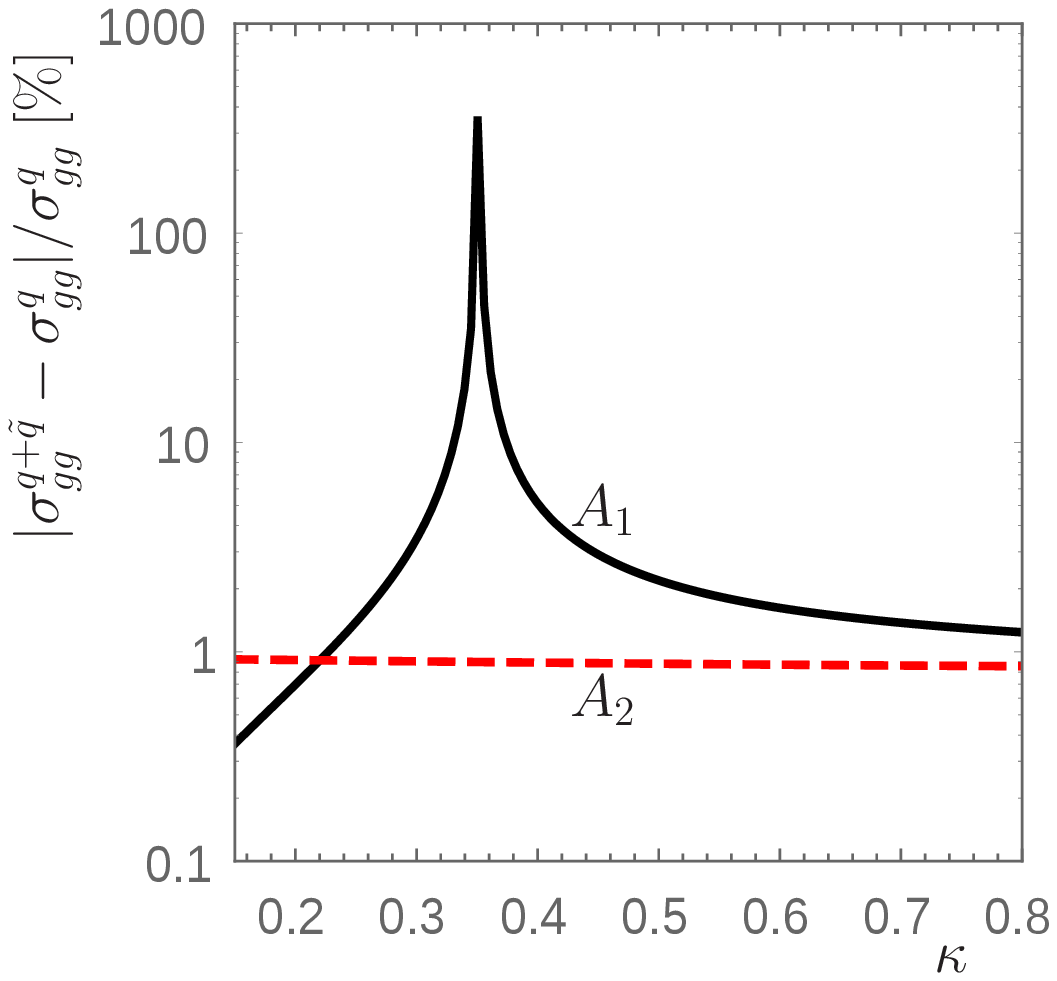} & 
\includegraphics[width=0.31\textwidth]{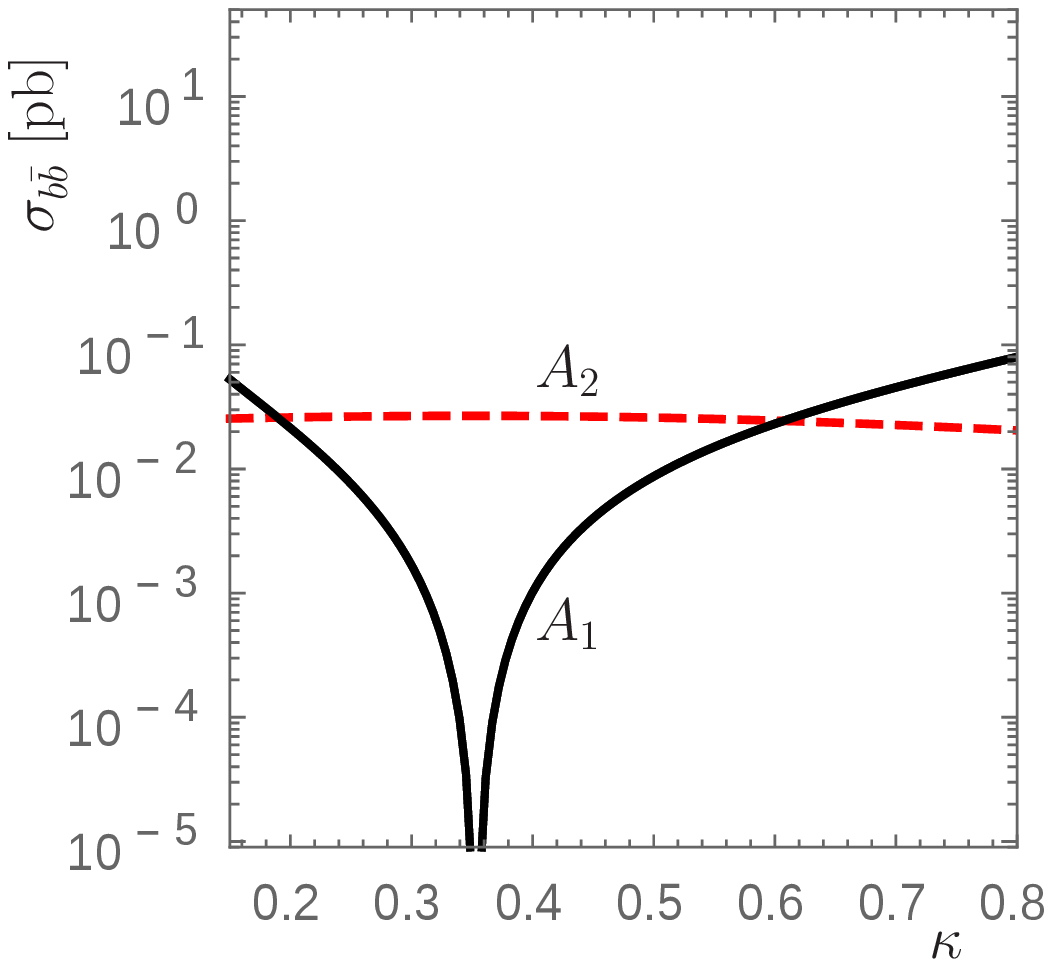}  \\[-0.4cm]
 (a)  & (b)  & (c) 
\end{tabular}
\end{center}
\vspace{-0.6cm}
\caption{(a) Gluon fusion and (c) bottom-quark annihilation in pb at $\sqrt{s}=13$\,TeV
as well as (b) absolute value of the relative squark corrections to gluon fusion
for the two \cp{}-odd Higgs bosons $A_1$ (black), $A_2$ (red, dashed)
as a function of $\kappa$ for scenario~$S_1$.}
\label{fig:CPoggh} 
\end{figure}

Similarly we depict the gluon fusion and bottom-quark annihilation
cross sections for the two \cp{}-odd Higgs bosons in \fig{fig:CPoggh}~(a)
and (c).
The pure singlet-like Higgs boson $A_1$ at $\kappa\sim 0.352$
is clearly apparent, since both cross sections vanish.
The corrections through squark contributions as shown in
\fig{fig:CPoggh}~(b) are very large around $\kappa\sim 0.352$,
since squark contributions are not suppressed through Higgs mixing,
although they only appear at \nlo{} \sqcd{}.
Electro-weak corrections induced through light quarks
are absent for the \cp{}-odd Higgs bosons.
We note that in the range $\kappa = 0.351-0.353$, where the \lo{} \qcd{}
gluon fusion cross section for the light \cp{}-odd Higgs boson $A_1$
are tiny, $<10^{-5}$\,pb, the prediction for $\sigma_{gg}^{q+\tilde{q}}$
with squark induced \nlo{} \sqcd{} contributions for $A_1$ is unreliable,
since \sushi{} calculates \nlo{} \qcd{} contributions through 
the multiplication of one-loop and two-loop contributions, where
the latter tend to be significantly larger than the former and can thus
even induce negative cross sections. However, in these regions the
tiny cross sections are not of relevance for current searches.
The cross section $\sigma_{gg}^{q+\tilde{q}}$ and the relative correction to
the vanishing only quark induced cross sections of more than $100$\% need
to be taken with care for the \cp{}-odd Higgs bosons.
The fact that for a pure singlet-like \cp{}-odd Higgs boson the
gluon fusion cross section at \nlo{} \sqcd{} completely vanishes due
to the absence of a \lo{} contribution motivates to take into
account \nnlo{} \sqcd{} stop contributions as it was done
for the light \cp{}-even Higgs boson in \citere{Bagnaschi:2014zla}.
A first estimate yields tiny, positive cross sections,
but we leave an inclusion in \sushi{} to future work.
The \cp{}-even Higgs bosons in contrast have a \lo{} squark induced contribution,
which leaves $\sigma_{gg}^{q+\tilde{q}}$ mostly well-behaved. 
Only in rare cases, where \lo{} squark and quark contributions cancel,
similar difficulties can arise.

\subsection{Scenario $S_2$: Inclusive cross sections for $\sqrt{s}=13$\,TeV}

\begin{figure}[htp]
\begin{center}
\begin{tabular}{ccc}
\includegraphics[width=0.31\textwidth]{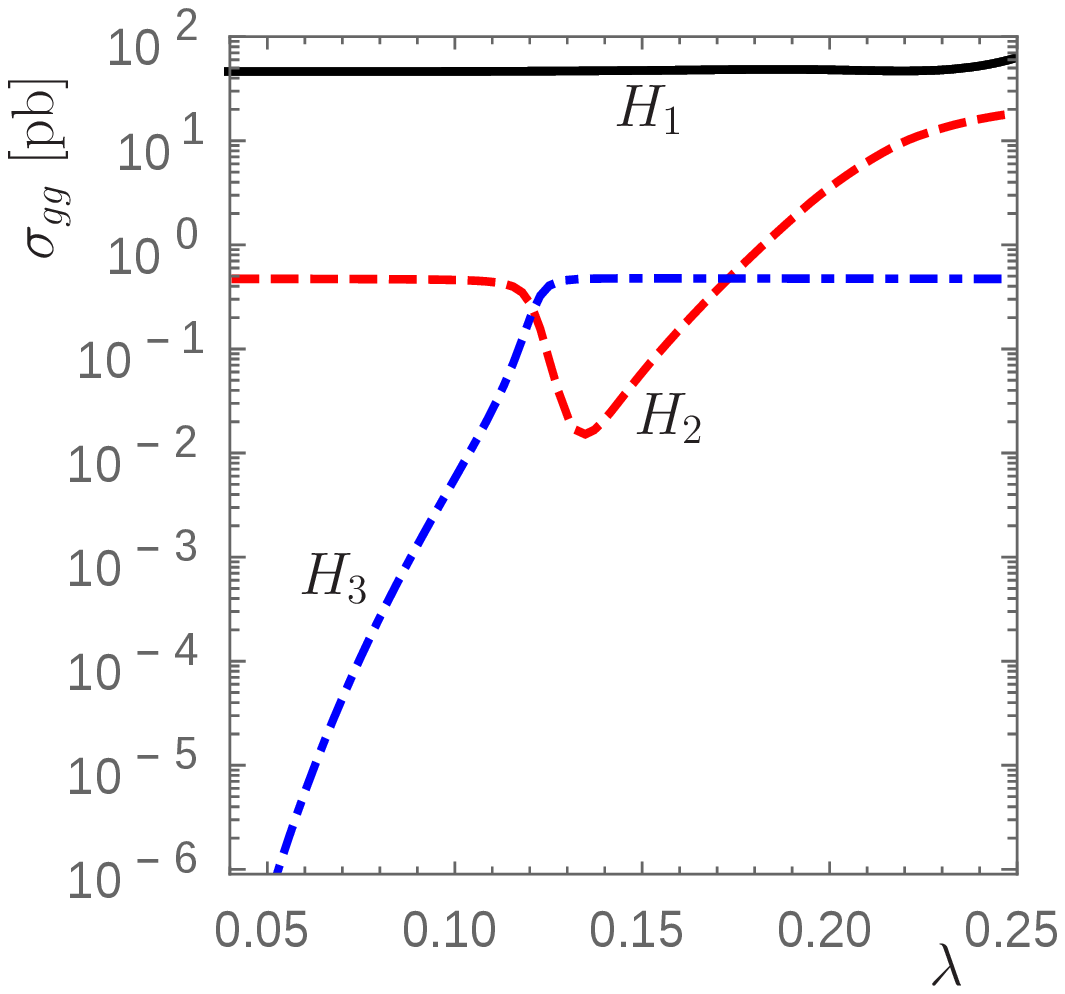} & 
\includegraphics[width=0.31\textwidth]{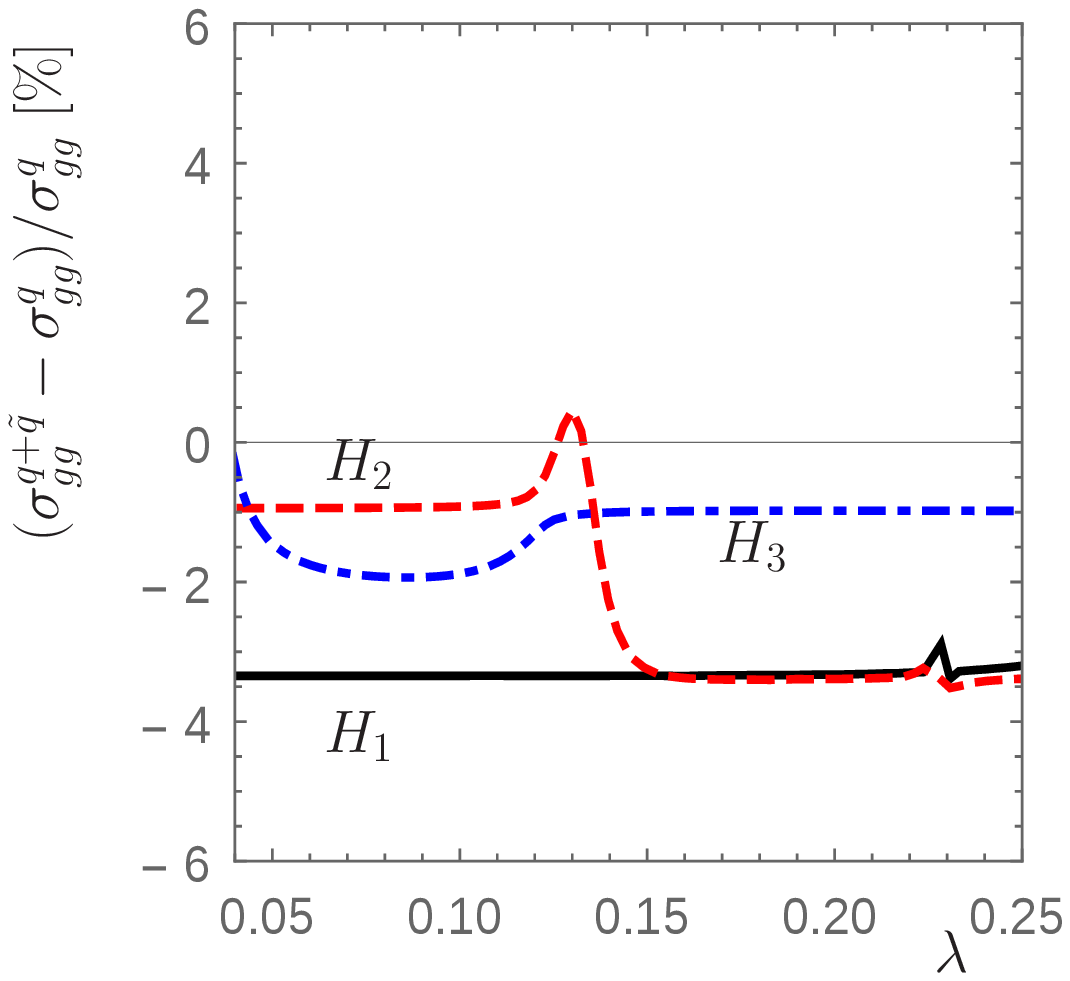} & 
\includegraphics[width=0.31\textwidth]{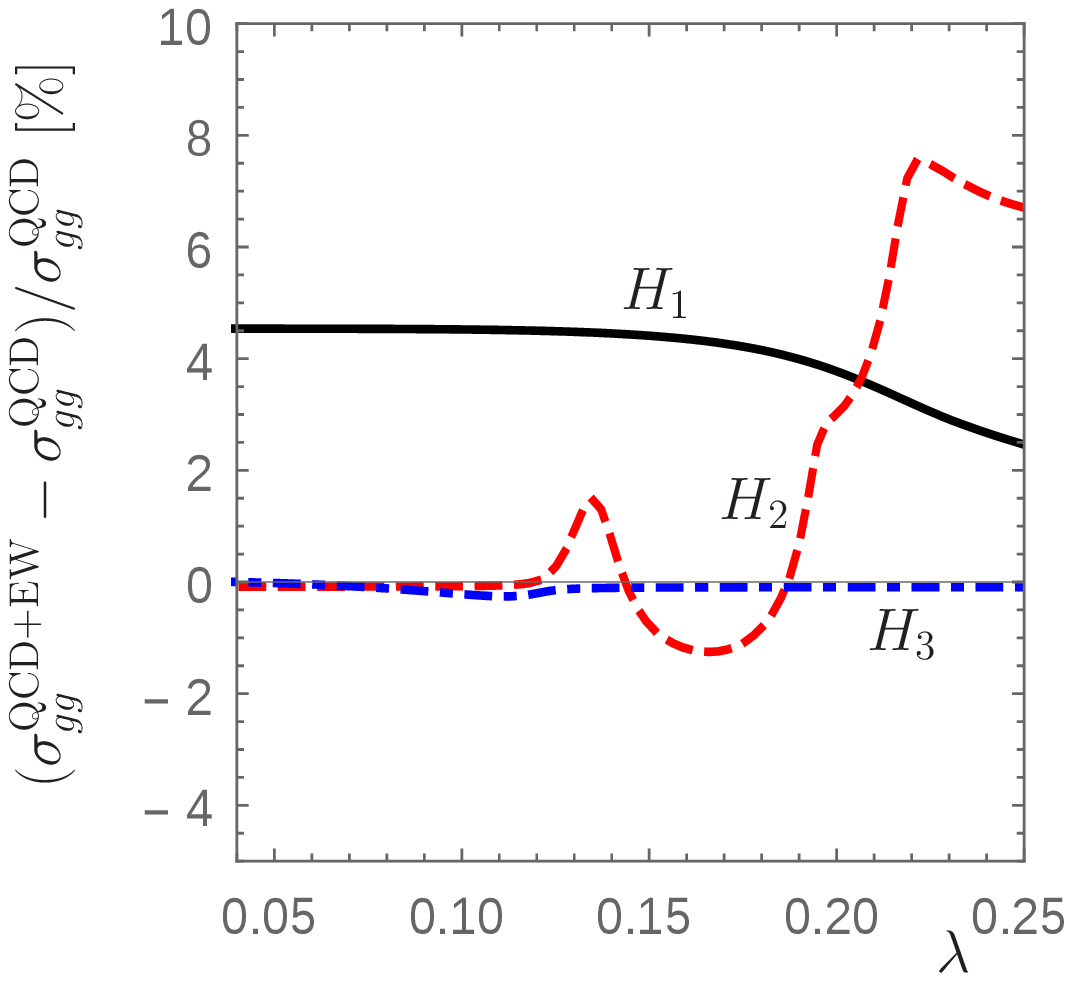}  \\[-0.4cm]
 (a) & (b) & (c)
\end{tabular}
\end{center}
\vspace{-0.6cm}
\caption{(a) Gluon fusion in pb at $\sqrt{s}=13$\,TeV, (b) squark corrections
and (c) electroweak corrections to gluon fusion
for the three \cp{}-even Higgs bosons $H_1$ (black), $H_2$ (red, dashed), $H_3$ (blue, dotdashed)
as a function of $\lambda$ for scenario~$S_2$.}
\label{fig:CPegghS6} 
\begin{center}
\begin{tabular}{ccc}
\includegraphics[width=0.31\textwidth]{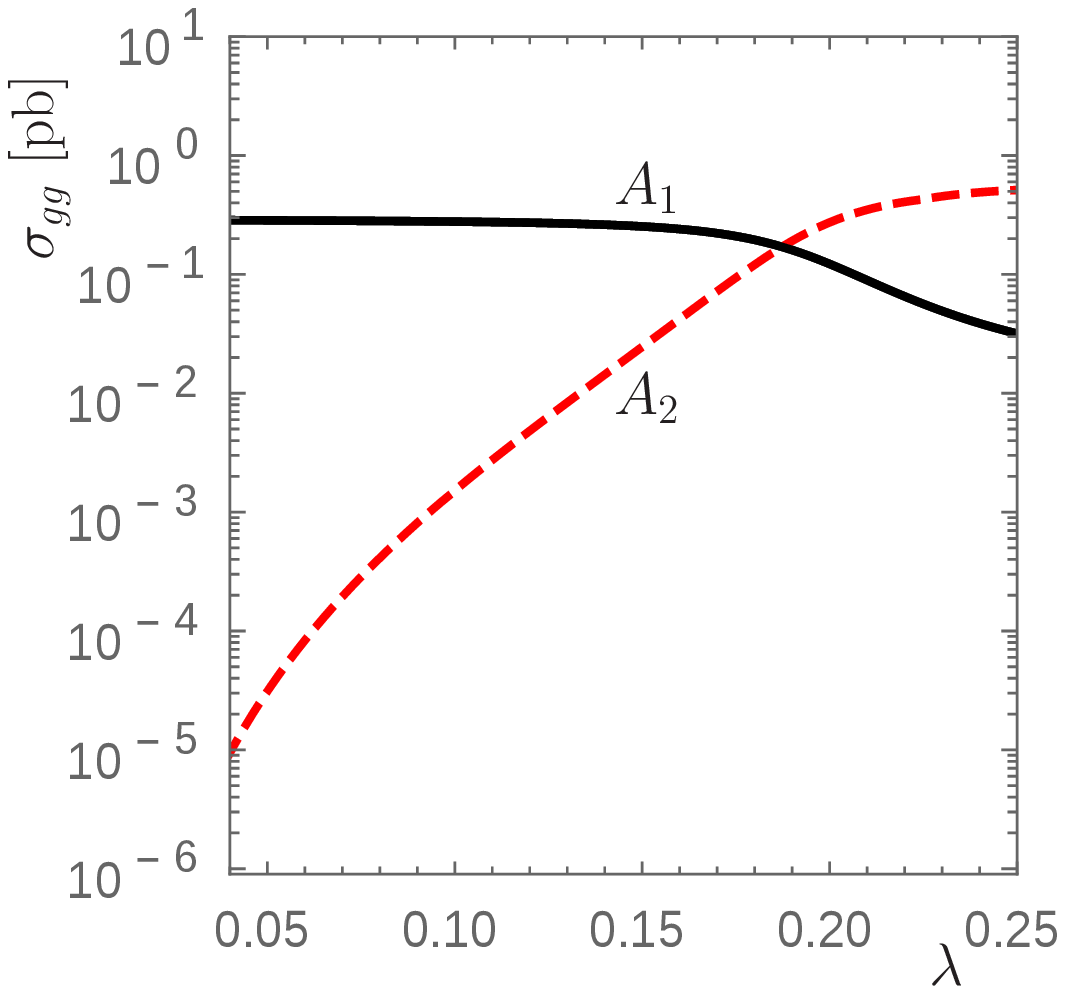} & 
\includegraphics[width=0.31\textwidth]{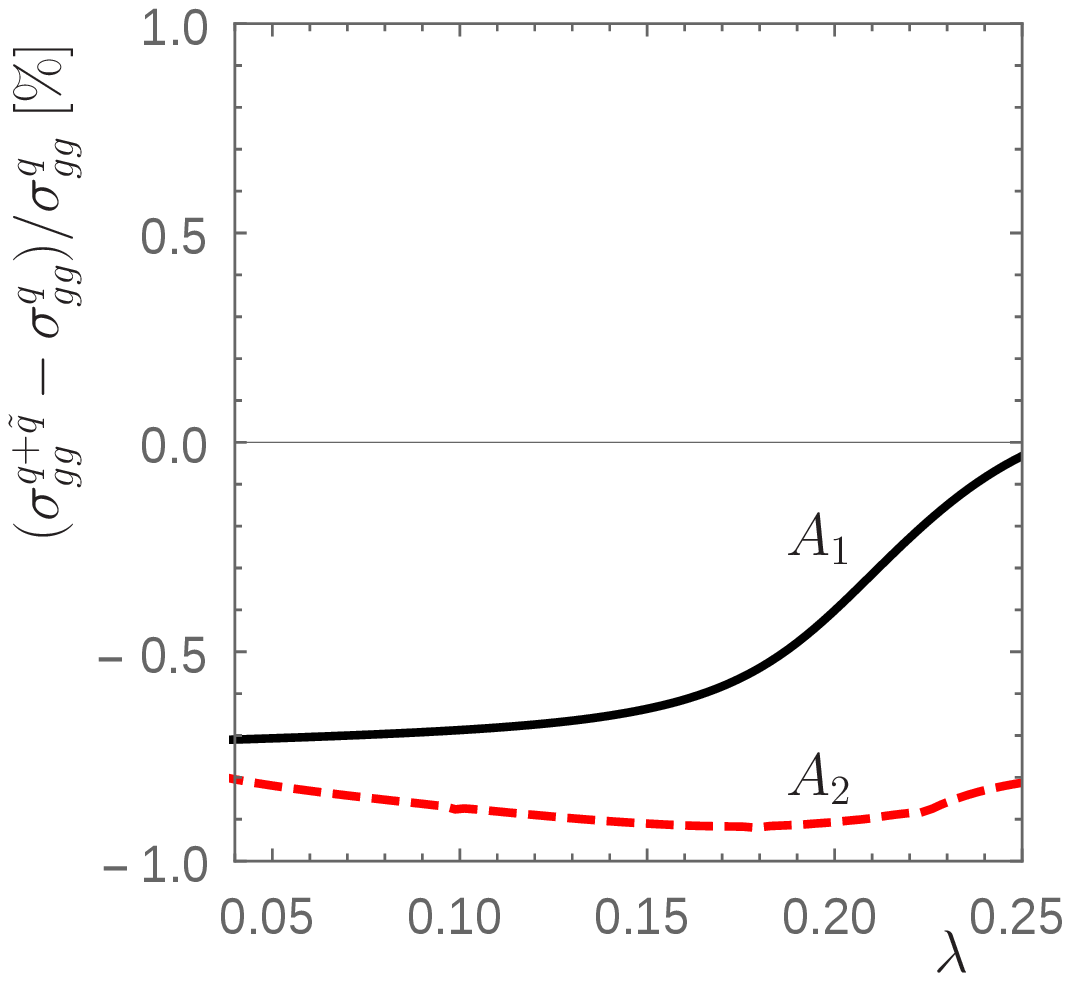} & 
\includegraphics[width=0.31\textwidth]{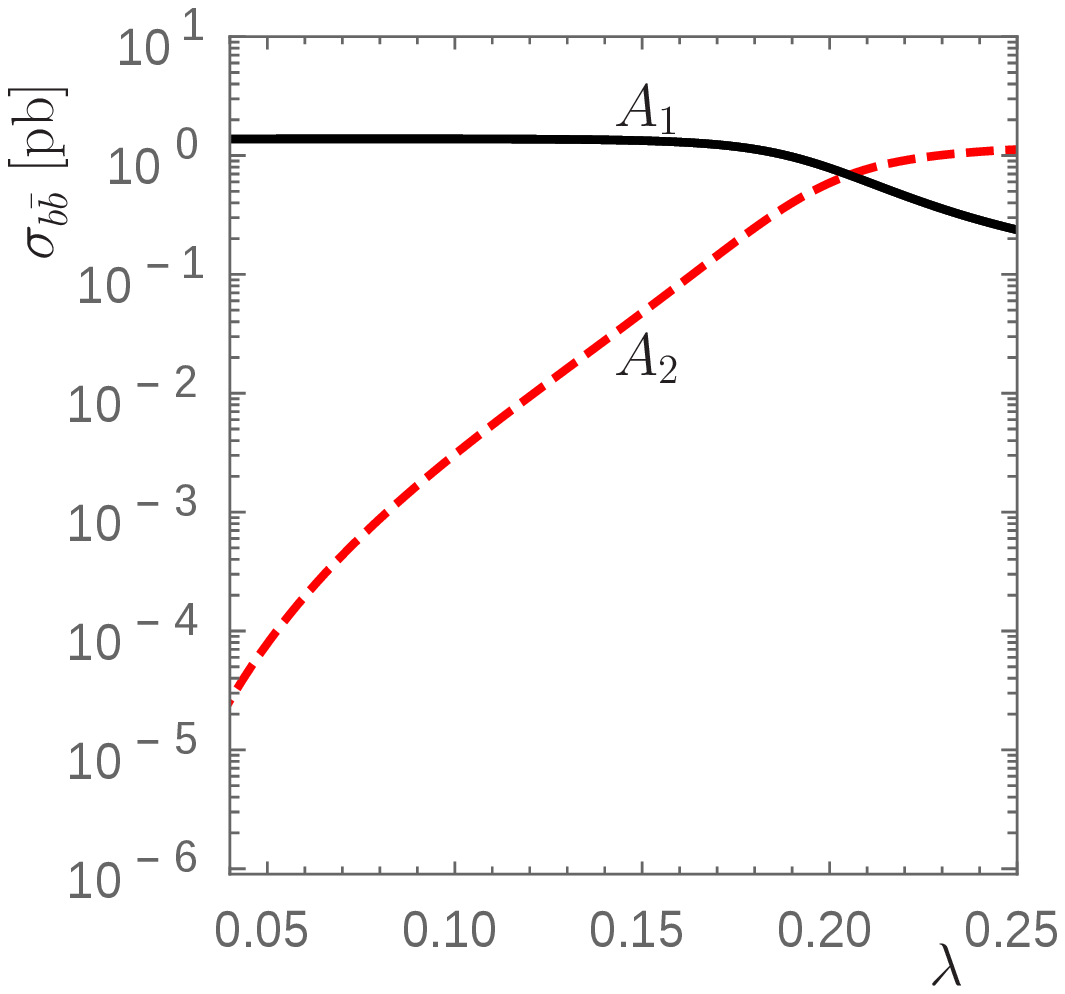}  \\[-0.4cm]
 (a) & (b)  & (c)
\end{tabular}
\end{center}
\vspace{-0.6cm}
\caption{(a) Gluon fusion and (c) bottom-quark annihilation in pb at $\sqrt{s}=13$\,TeV
as well as (b) the relative squark corrections to gluon fusion
for the two \cp{}-odd Higgs bosons $A_1$ (black), $A_2$ (red, dashed)
as a function of $\lambda$ for scenario~$S_2$.}
\label{fig:CPogghS6} 
\end{figure}

In this subsection we examine the inclusive cross sections for scenario $S_2$,
which includes low values of $\lambda$ and thus reflects the decoupling
of the singlet-like Higgs. The gluon fusion cross section together with
squark contributions and electroweak corrections through light quarks are
shown in \fig{fig:CPegghS6} for the three \cp{}-even Higgs bosons.
With decreasing $\lambda$ the gluon fusion cross section for $H_{2/3}$ rapidly decreases,
which has two reasons: First the increasing mass naturally decreases the
cross section, but second the singlet-like Higgs boson also decouples from
the other two Higgs bosons -- as apparent in \fig{fig:CPodefine}.
Thus, the indirect coupling to quarks is suppressed. Moreover also
the direct couplings to squarks are proportional to~$\lambda$ and thus
decrease in size with decreasing $\lambda$. The relative
correction induced by squark contributions remains rather constant,
see \fig{fig:CPegghS6}~(b). The small interference structure
visible around $\lambda=0.22-0.23$ stems from the interchange of the dominant
singlet character between $H_1$ and $H_2$.
Remark that for a \sm{} Higgs the decrease of the gluon fusion cross
section due to the increase in mass is between $m_H=500$\,GeV
and $m_H=1200$\,GeV only a factor of~$\sim 47$, whereas we observe
a decrease of more than five orders of magnitude, thus mainly driven by the
decoupling. Electro-weak corrections by light quarks as depicted
in \fig{fig:CPegghS6}~(c) are completely absent for a heavy
singlet-like Higgs boson $H_3$, but show a similar pattern at large
$\lambda$ for $H_2$ as for small $\kappa$ in scenario $S_1$.
The reason is that the Higgs mass $m_{H_2}$ in both scenarios crosses
the $2m_Z$ and $2m_W$ thresholds for the electroweak corrections
by light quarks.
The bottom-quark annihilation cross section shows a similar decoupling
behavior and is thus not explicitly shown.

For the \cp{}-odd Higgs bosons we show the corresponding decoupling
limit in \fig{fig:CPogghS6}, where the gluon fusion cross section
and the bottom-quark annihilation cross section presented in
\fig{fig:CPogghS6}~(a) and (c) respectively decrease dramatically
for the singlet-like Higgs bosons $A_2$. Again the squark corrections shown
in \fig{fig:CPogghS6}~(b) remain rather constant.
Given the fact that the total cross section for $A_2$ vanishes
with decreasing $\lambda$, we can therefore conclude that both
the indirect couplings to quarks for $A_2$ vanish, but also
the direct couplings to squarks vanish. Thus, the singlet-like Higgs boson~$A_2$
decouples in the limit $\lambda$ being small.

We point out that the singlet-like Higgs bosons~$H_3$ and $A_2$ both
approach \susy{} mass thresholds with decreasing $\lambda$. Therefore
we employ the lower bound of $\lambda=0.04$, since for lower values
of $\lambda$ and thus larger masses of $H_3$ and $A_2$ we cannot
guarantee the validity of the \nlo{} \sqcd{} contributions implemented
in \sushi{}. \citere{Bagnaschi:2014zla} therefore assigned an additional
theoretical uncertainty to the heavy \susy{} masses expansion.
In the decoupling regime we checked that the cross sections for $H_1$, $H_2$
and $A_1$ coincide with the \mssm{} cross sections obtained for a
mixing angle of $\alpha=-0.12347$ with an accuracy of $\sim 10^{-4}$, which
resembles the remaining singlet fraction of $H_1$, $H_2$ and $A_1$.

\subsection{Theory uncertainties}

In this section we shortly focus on theoretical uncertainties in the calculation
of neutral Higgs boson production cross sections. \citere{Bagnaschi:2014zla} identified the most
important theoretical uncertainties for the \mssm{}, which mostly apply to our discussion
of the \nmssm{} as well. Apart from the well-known renormalization and
factorization scale and PDF$+\alpha_s$ uncertainties
for cross sections at a proton-proton collider an additional uncertainty
for gluon fusion cross section is
the choice of a renormalization scheme for the bottom-quark Yukawa coupling, which
is of particular relevance if the bottom-quark loop dominantly contributes.
Secondly, the fact that \nlo{} \sqcd{} contributions are taken into
account in an expansion of heavy \susy{} masses induces an uncertainty,
which grows for larger Higgs masses approaching \susy{} particle
masses thresholds. Thirdly, also relevant for bottom-quark annihilation are
missing contributions in the resummation~$\Delta_b$, which
induce an uncertainty, in particular in the limit $\Delta_b\rightarrow -1$.
All of the above theoretical uncertainties as discussed in \citere{Bagnaschi:2014zla}
apply to the \nmssm{} in a similar way. In contrast to the \mssm{} however
phenomenological studies of the \nmssm{} focus on lower values of $\tan\beta$,
where both the uncertainty from the choice of the bottom-quark Yukawa coupling and the
uncertainty induced from unknown contributions to $\Delta_b$ are of less importance.
A detailed discussion
in particular for the singlet-like \cp{}-even and \cp{}-odd Higgs boson is 
left for future work.

In the following we stick to the commonly studied renormalization
and factorization scale uncertainties as well as the PDF$+\alpha_s$ uncertainties.
We present our results just for scenario~$S_1$,
since no generically new features appear in other \susy{} scenarios.
We start with the scale uncertainty, where
we follow the prescription employed in \citeres{Dittmaier:2011ti,Bagnaschi:2014zla}.
We thus consider seven combinations of renormalization and
factorization scales defined as set $C_\mu$ of pairs $(\muR,\muF)$ with
$\muR=\left\lbrace m_\phi/4,\,m_\phi/2,\,m_\phi\right\rbrace$ and
$\muF=\left\lbrace m_\phi/4,\,m_\phi/2,\,m_\phi\right\rbrace$ under the
constraint $1/2 \leq \muR/\muF \leq 2$ for gluon fusion.
For bottom-quark annihilation the set is determined from
$\muR=\left\lbrace m_\phi/2,m_\phi,2\,m_\phi \right\rbrace$ and
$\muF=\left\lbrace m_\phi/8,m_\phi/4,m_\phi/2\right\rbrace$ with the constraint
$2 \leq \muR/\muF \leq 8$.
The minimal and maximal cross sections are obtained according to
\begin{align}
\label{eq:sigmapm}
\sigma^-:= \min_{(\muR,\,\muF)\,\in \,C_\mu} \left\{\sigma(\muR,\muF)\right\}\quad,\qquad
\sigma^+:= \max_{(\muR,\,\muF)\,\in\, C_\mu} \left\{\sigma(\muR,\muF)\right\}\quad,
\end{align}
which we present relative to the cross sections $\sigma(\muR^0,\muF^0)$
at the central scales $\muR^0$ and $\muF^0$. They are
$\muR^0=\muF^0=m_\phi/2$ for gluon fusion and
$\muR^0=m_\phi$ and $\muF^0=m_\phi/4$ for bottom-quark annihilation.
We therefore define the relative uncertainties
\begin{align}
\label{eq:deltamu}
\Delta^+_\mu :=
\frac{\sigma^+-\sigma(\muR^0,\muF^0)}{\sigma(\muR^0,\muF^0)}\quad,
\qquad
\Delta^-_\mu :=
\frac{\sigma^--\sigma(\muR^0,\muF^0)}{\sigma(\muR^0,\muF^0)}\quad.
\end{align}

\begin{figure}[htp]
\begin{center}
\begin{tabular}{cc}
\includegraphics[width=0.47\textwidth]{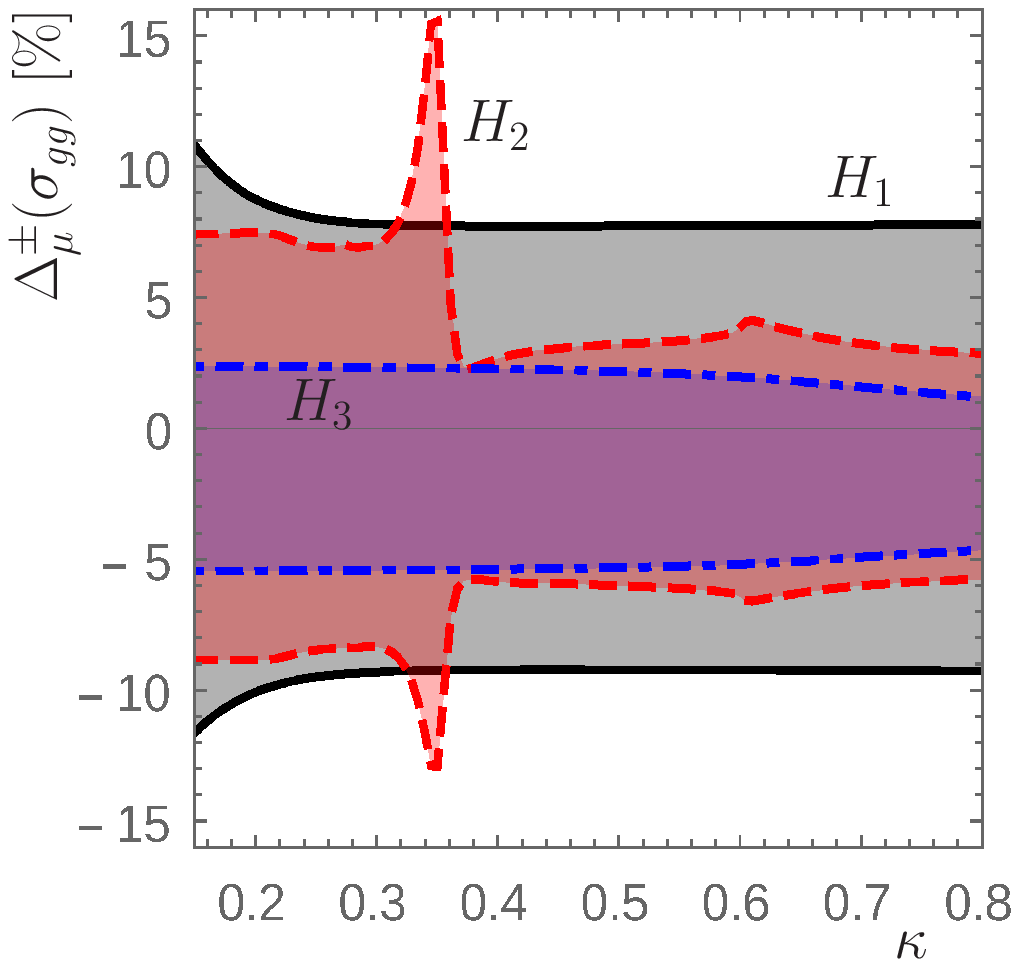} & 
\includegraphics[width=0.47\textwidth]{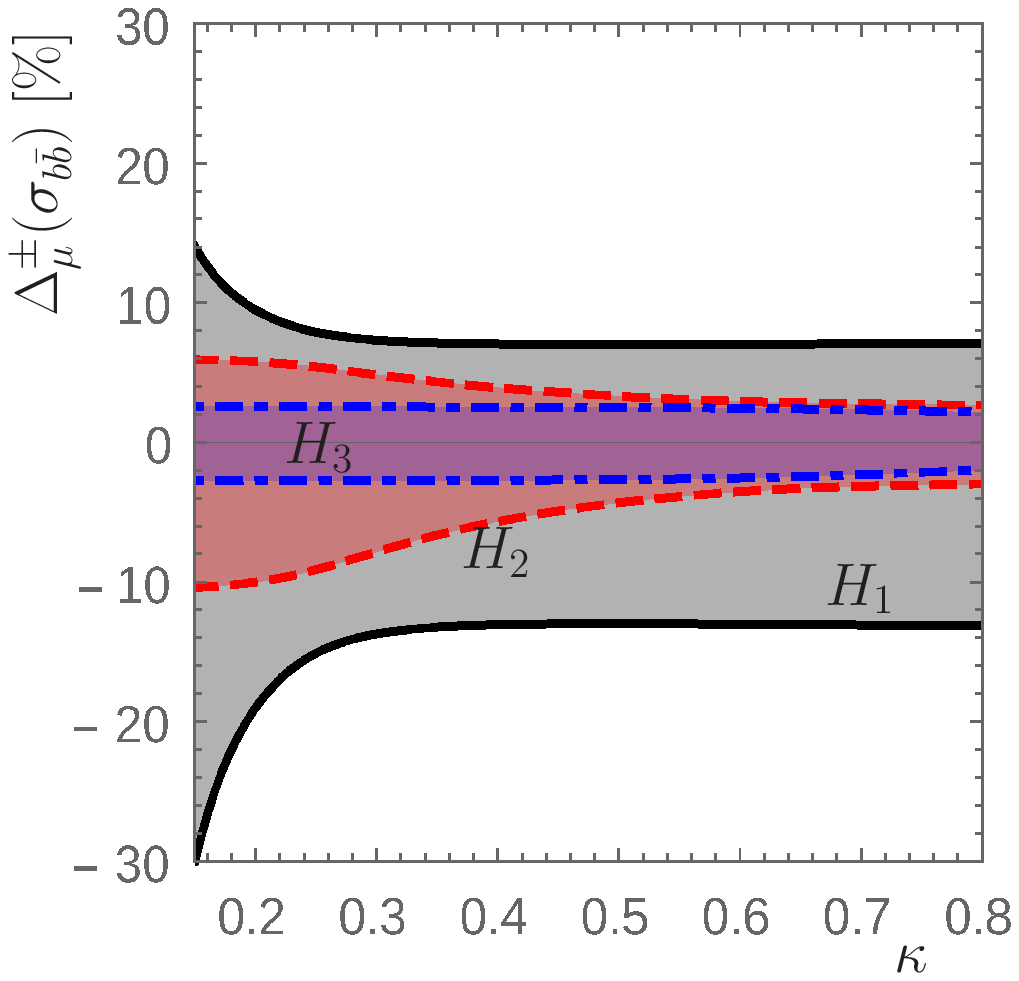}  \\[-0.4cm]
 (a)  & (b)\\
\includegraphics[width=0.47\textwidth]{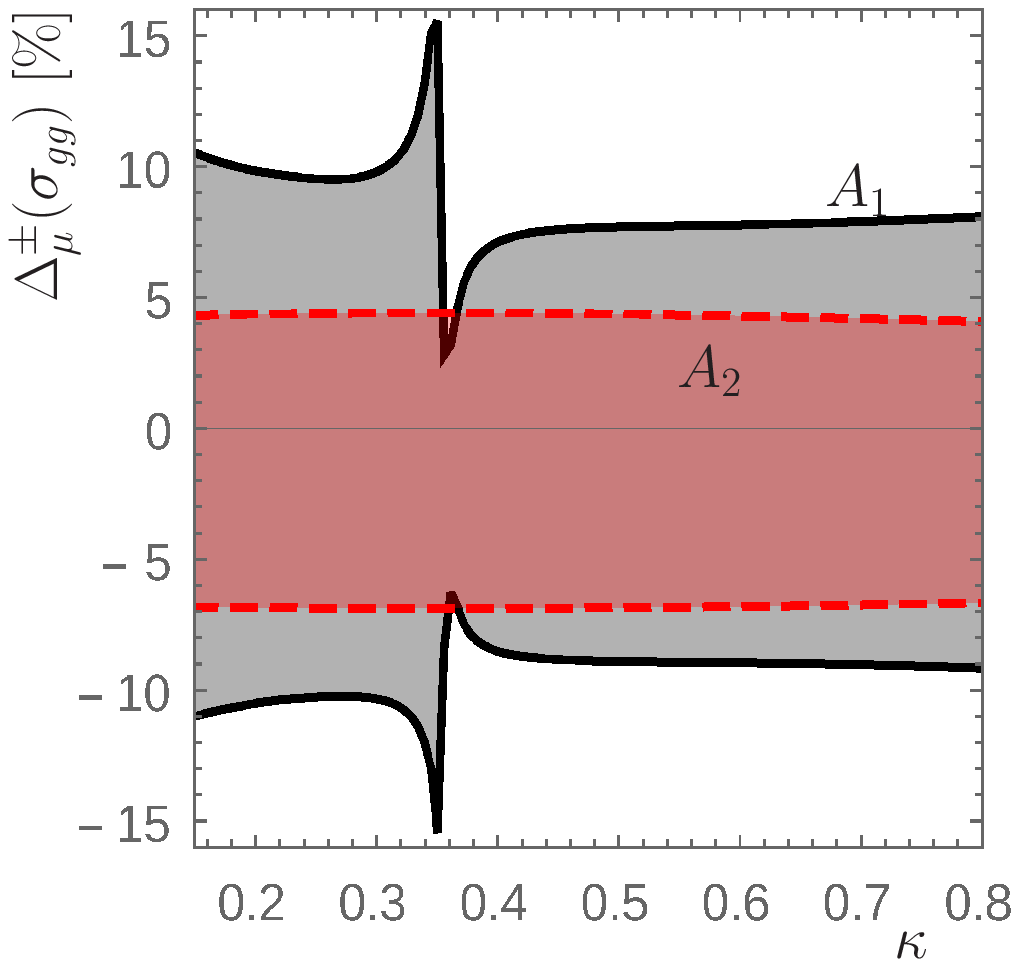} & 
\includegraphics[width=0.47\textwidth]{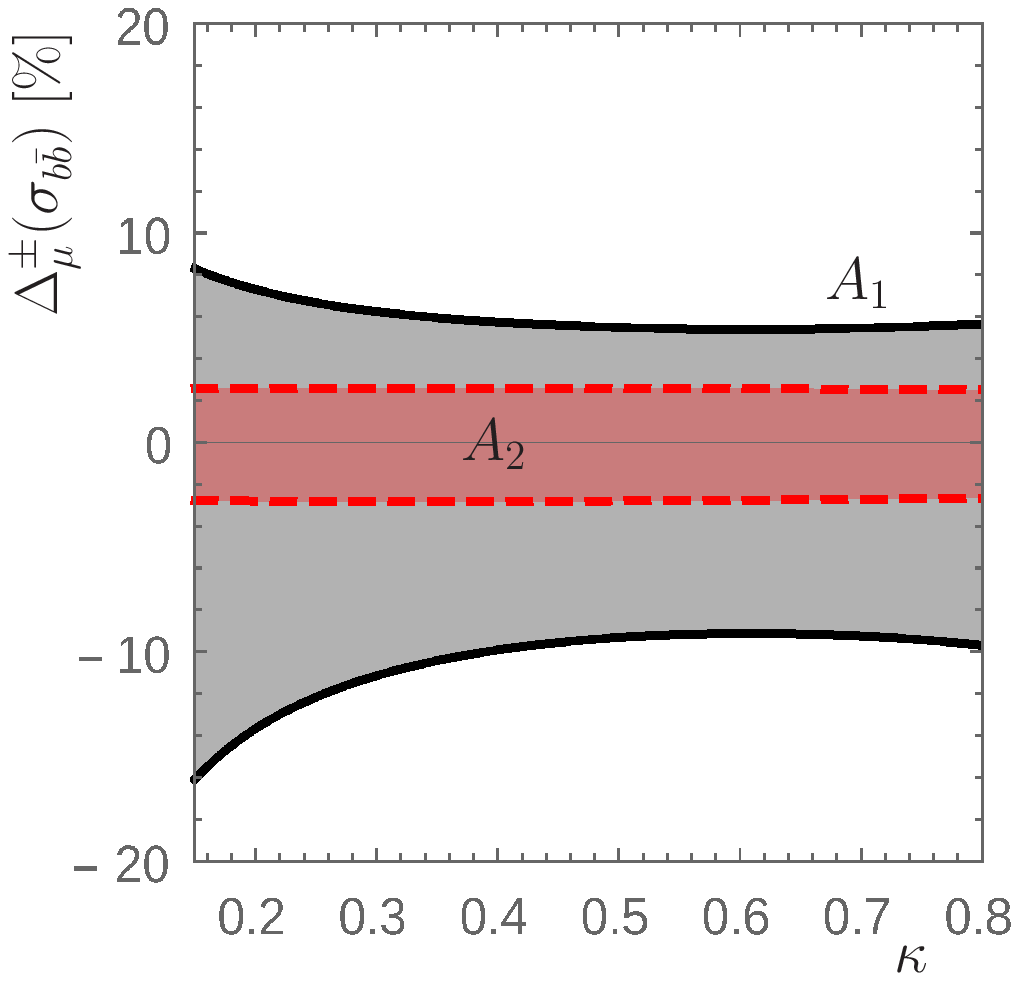}  \\[-0.4cm]
 (c) & (d) 
\end{tabular}
\end{center}
\vspace{-0.6cm}
\caption{Scale uncertainties $\Delta_\mu^+$ and $\Delta_\mu^-$
for (a,c) gluon fusion and (b,d) bottom-quark annihilation for
(a,b) the three \cp{}-even Higgs bosons
$H_1$ (black), $H_2$ (red, dashed), $H_3$ (blue, dotdashed)
and for (c,d) the two \cp{}-odd Higgs bosons $A_1$ (black), $A_2$ (red, dashed)
as a function of $\kappa$ for scenario~$S_1$.
}
\label{fig:CPscales} 
\end{figure}

The scale uncertainties are shown for both the \cp{}-even
and \cp{}-odd Higgs bosons in \fig{fig:CPscales} for $\sqrt{s}=13$\,TeV.
In case of gluon fusion
the \sm{}-like Higgs comes with a scale uncertainty
of about $\mathcal{O}(\pm 10$\%) taking into account \nnlo{} \qcd{} top-quark
contributions in the heavy top-quark effective theory.
The scale uncertainty is naturally strongly dependent
on the individual contributions to the cross section and increases
in particular in regions, where the top-quark induced contributions
are small or quark contributions to the gluon fusion cross section cancel
and come along with large squark
and/or electroweak corrections. The latter effect is very pronounced
for $H_2$ in \fig{fig:CPscales}~(a) and for $A_1$
in \fig{fig:CPscales}~(c).

For bottom-quark annihilation as shown in \fig{fig:CPscales}~(b) and (d)
the scale uncertainty is mainly
dependent on the Higgs mass, rather than the specific \susy{} scenario.
The large uncertainty for low Higgs masses reflects the need to move toward the
four-flavor scheme (4FS)~\cite{Dittmaier:2003ej,Dawson:2003kb} in the description of the process.

\begin{figure}[htp]
\begin{center}
\begin{tabular}{cc}
\includegraphics[width=0.47\textwidth]{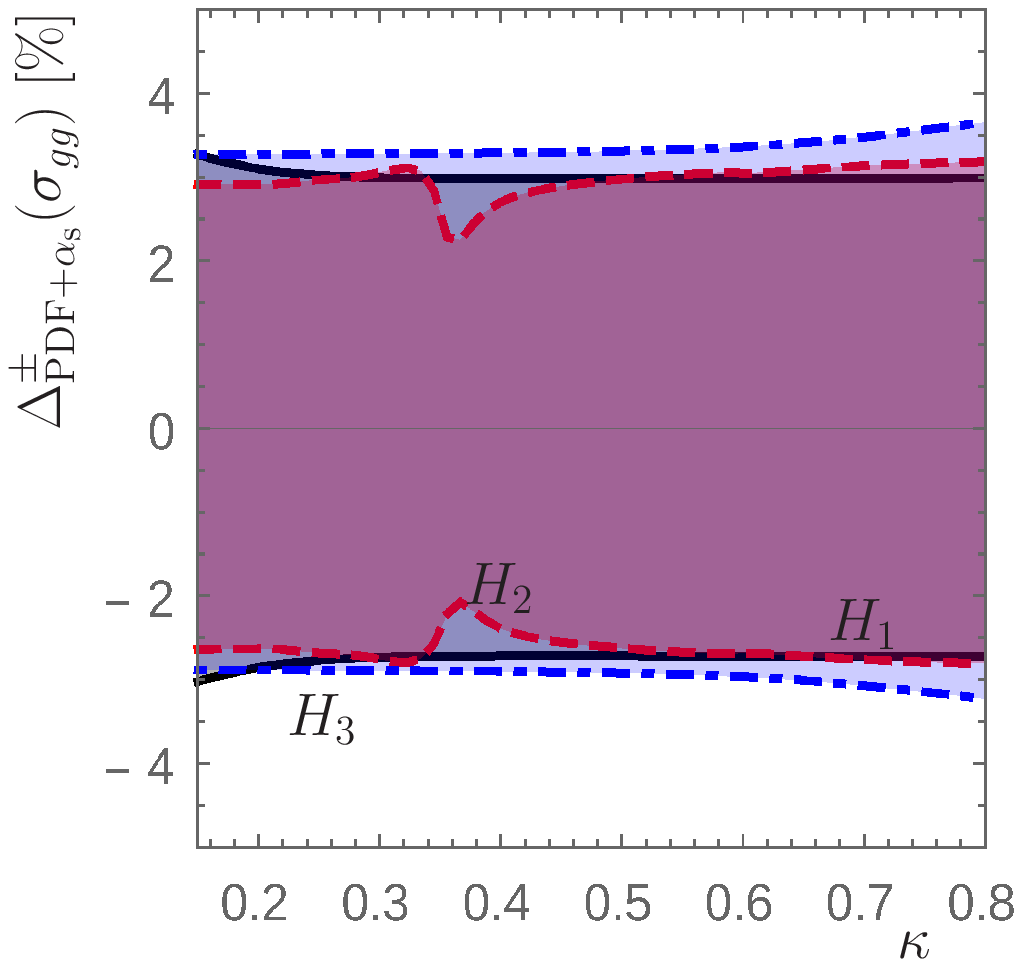} & 
\includegraphics[width=0.47\textwidth]{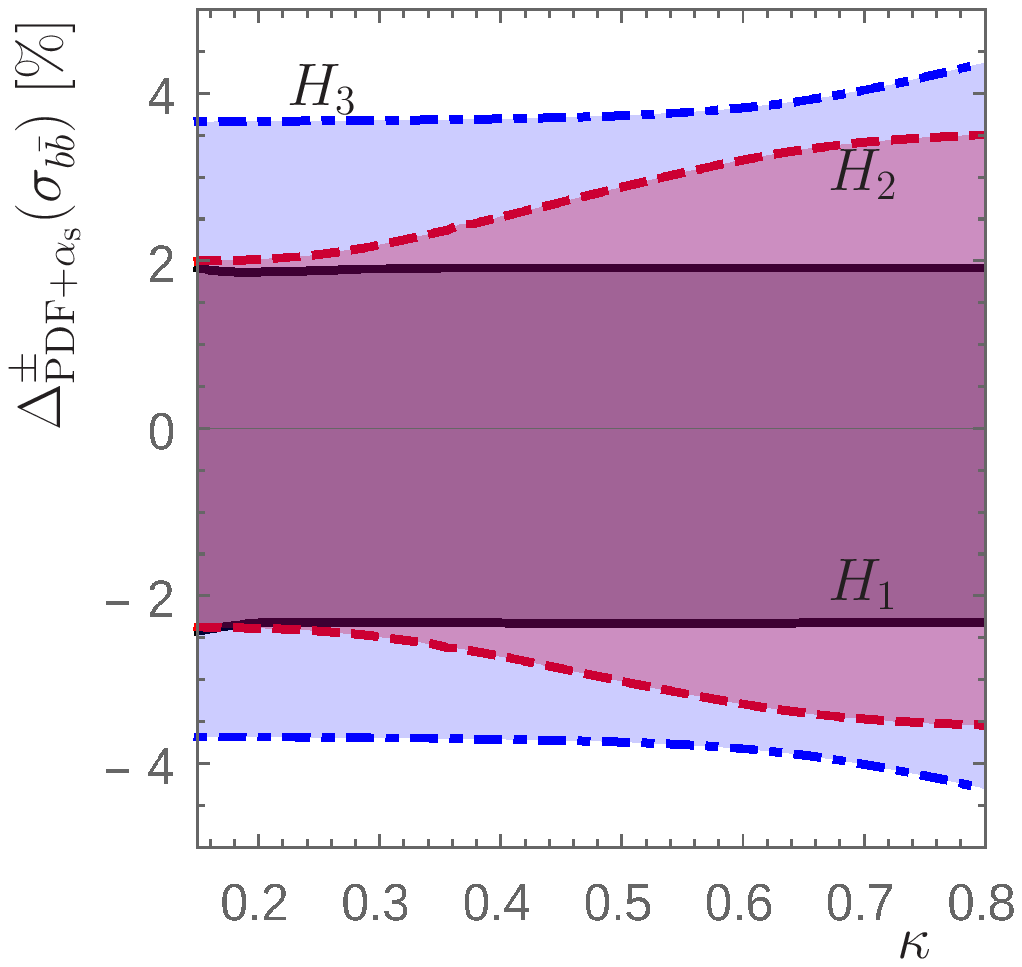}  \\[-0.4cm]
 (a)  & (b)
\end{tabular}
\end{center}
\vspace{-0.6cm}
\caption{PDF$+\alpha_s$ uncertainties
for (a) gluon fusion and (b) bottom-quark annihilation for
the three \cp{}-even Higgs bosons
$H_1$ (black), $H_2$ (red, dashed), $H_3$ (blue, dotdashed)
as a function of $\kappa$ for scenario $S_1$.}
\label{fig:CPpdfalphas} 
\end{figure}

As a last step we discuss PDF$+\alpha_s$ uncertainties by applying the 
practical PDF4LHC recommendation~\cite{Alekhin:2011sk,Botje:2011sn}
for the {\tt MSTW2008}~\cite{Martin:2009iq}
PDF sets in order to emphasize the findings of \citere{Bagnaschi:2014zla}
for the case of the \nmssm{}. For this purpose we combine the results obtained
with the $41$ PDF sets of {\tt MSTW2008(n)nlo68cl}
with the $\alpha_s$ uncertainties obtained by the PDF sets, which
vary $\alpha_s$ within the $68$\% confidence level interval.
\fig{fig:CPpdfalphas} shows the
PDF$+\alpha_s$ relative uncertainties $\Delta_{\rm{PDF}+\alpha_s}^{\pm}$ with respect
to the standard PDF$+\alpha_s$ choice for the three \cp{}-even Higgs bosons
for gluon fusion and bottom-quark annihilation. The standard PDF$+\alpha_s$ choice
equals the zeroth PDF set of {\tt MSTW2008(n)nlo68cl} together with the standard values
$\alpha_s=0.120$ at \nlo{} and $\alpha_s=0.117$ at \nnlo{} \qcd{}.
Similar to the \mssm{}
the uncertainties are mainly dependent on the Higgs mass and only slightly
dependent on the specific \susy{} scenario, even for the singlet-like Higgs boson.
A very similar result applies to the \cp{}-odd Higgs sector and is thus
not explicitly presented. It therefore
seems sufficient to take over the full relative PDF$+\alpha_s$ uncertainties from
a \cp{}-even or \cp{}-odd \sm{} Higgs boson with the same mass, which is easily adjustable
to future updates of the PDF4LHC recommendation.
Taking into account the combination of the newest
{\tt MMHT2014}~\cite{Harland-Lang:2014zoa},
{\tt NNPDF 3.0}~\cite{Ball:2014uwa}
and {\tt CT10}~\cite{Lai:2010vv} PDF sets naturally results
in larger PDF$+\alpha_s$ uncertainties. However, the simple recipe to obtain PDF$+\alpha_s$
uncertainties just as a function of the Higgs mass is applicable
for the combination of the PDF sets provided by the PDF fitting groups as well.

\newpage
\section{Conclusions}
\label{sec:conclusions}

We presented accurate predictions for neutral Higgs boson production 
at proton colliders through
gluon fusion and bottom-quark annihilation in the \cp{}-conserving \nmssm{}.
For gluon fusion we adapt the full \nlo{} \qcd{}
and \sqcd{} results from the \mssm{} to the \nmssm{},
based on an asymptotic expansion in
heavy \susy{} masses for squark and squark/quark/gluino two-loop
contributions. Top-quark induced \nnlo{} \qcd{} contributions
are added in the heavy top-quark effective theory.
Electro-weak corrections to gluon fusion mediated through light quarks
are taken into account and the resummation of sbottom
contributions for large values of $\tan\beta$ can be translated from
the \mssm{} to the \nmssm{}. The latter procedure also applies to bottom-quark annihilation.

Our discussion comes along with an implementation of the neutral Higgs boson production
cross section calculation in the code \sushi{}. The Higgs sector (obtained
by an \nmssm{} spectrum generator) needs to be supplied through
the \sushi{} input file.
We briefly focused on the new features of the additional singlet-like
\cp{}-even or \cp{}-odd Higgs boson
for what concerns neutral Higgs boson production. Due to possible cancellations
of quark induced contributions, squark and electroweak corrections
to gluon fusion can be of greater relevance than known in the \mssm{}, in particular
for not too heavy third generation squark mass spectra.
For a small singlet-doublet mixing term, which can be achieved
by lowering the parameter~$\lambda$,
the singlet-like \cp{}-even and -odd Higgs boson can both be decoupled
from the remaining \mssm{}-like Higgs sector.
The renormalization and factorization scale uncertainties reflect
the individual contributions to neutral Higgs boson production in case
of gluon fusion, whereas scale uncertainties for bottom-quark annihilation
as well as PDF$+\alpha_s$ uncertainties for both production processes
mainly remain a function of the Higgs boson mass.

We leave a more detailed investigation of theory uncertainties to future
work. Moreover interesting for future studies is an expansion in
a light Higgs boson mass rather than heavy \susy{} masses
for what concerns the inclusion of \nlo{}
and \nnlo{} \sqcd{} contributions, in particular since
for pure singlet-like \cp{}-odd Higgs bosons \nnlo{} stop-induced
contributions are the first non-vanishing
contributions to gluon fusion.
Similarly a discussion of distributions and of the necessity of
resummation for transverse momentum distributions is timely
for the real \nmssm{}, but left for future work.

\section*{Acknowledgments}
The author thanks Jonathan Gaunt, Robert Harlander, Hendrik Mantler and
Pietro Slavich for very helpful comments on the manuscript.
The author is, moreover, indebted to Pietro Slavich for help in the translation
of the \mssm{} \nlo{} \sqcd{} corrections to the \nmssm{} and to Robert Harlander
and Hendrik Mantler for their comments on the implementation
of the \nmssm{} in \sushi{}. The author also thanks Kathrin Walz for help related to the renormalization
of the stop and sbottom sectors within {\tt NMSSMCALC} and Peter Drechsel for help
with regard to the calculation of \nmssm{} Higgs boson masses.
The author acknowledges support by Deutsche Forschungsgemeinschaft (DFG) through the 
Collaborative Research Center~SFB~676~``Particles, Strings and the Early Universe''.

\appendix
\section{Formulas: Higgs-squark-squark couplings in the \nmssm{}}
\label{app:higgssquark}

In this section we present the squark couplings to the five neutral Higgs bosons $\phi$ of the \nmssm{}
as implemented in the code \sushi{}. As noted before, the singlet component $S$ does not couple to
quarks, such that the couplings of the Higgs bosons to quarks can be taken over from the \mssm{}
by replacing the mixing angle $\alpha$ and thus the projection on $H_d^0$ and $H_u^0$
by the proper mixing matrix elements $\RS/\RP$. On the contrary the singlet component couples to
squarks, for which we present the Feynman rules in the form
\begin{align}
\parbox{27mm}{\includegraphics[width=0.2\textwidth]{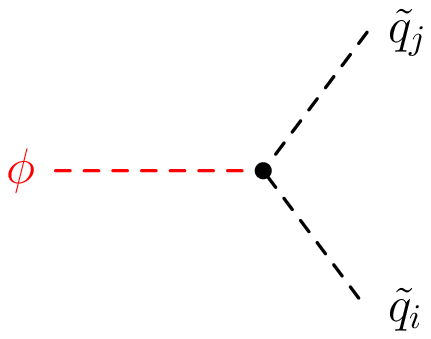}}=i\frac{m_{q}^2}{v}g_{\tilde{q},ij}^{\phi}\quad,
\end{align}
with $v = 1/\sqrt{\sqrt{2} G_F}= \sqrt{v_d^2+v_u^2}$. 
The couplings $g_{\tilde{q},ij}^{\phi}$ of squarks
with indices~$\lbrace i,j\rbrace$ to \cp{}-even Higgs
bosons with $k=\lbrace 1,2,3\rbrace$ or \cp{}-odd Higgs
bosons with $k=\lbrace 1,2\rbrace$ are subsequently
presented in gauge eigenstates
\begin{align}
&g_{\tilde{q},ij}^{H_k} = \RS_{k 1}\tilde{g}_{\tilde{q},ij}^{H1}
 + \RS_{k 2}\tilde{g}_{\tilde{q},ij}^{H2}
 + \RS_{k 3}\tilde{g}_{\tilde{q},ij}^{H3}\\
&g_{\tilde{q},ij}^{A_k} = \RP_{k1}\cdot(c_\beta \tilde{g}_{\tilde{q},ij}^{A1} - s_\beta \tilde{g}_{\tilde{q},ij}^{A2})
 + \RP_{k2}\cdot (s_\beta \tilde{g}_{\tilde{q},ij}^{A1} + c_\beta \tilde{g}_{\tilde{q},ij}^{A2})
 + \RP_{k3}\tilde{g}_{\tilde{q},ij}^{A3}\quad,
\end{align}
where in the \cp{}-odd sector the prerotation with
$\RG$ involving $c_\beta=\cos\beta$ and $s_\beta=\sin\beta$ is performed.
They were obtained with the code {\tt MaCoR}~\cite{Liebler:2010bi} and cross-checked
against the formulas of \citere{King:2012tr} for what concerns the \lo{} stop contributions
to gluon fusion.
The individual contributions $\tilde{g}_{\tilde{q},ij}^{Hk}$
in the \cp{}-even sector with $k=\lbrace 1,2,3\rbrace$ yield:

{\allowdisplaybreaks
\begin{align}
m_b^2\tilde{g}_{\tilde{b},11}^{H1} &= 
      \frac{2 m_b^2}{c_\beta} - \frac{1}{6}m_Z^2 c_\beta\left[3 + \cbt (1+2\ctw)\right]
      + \frac{\sbt}{2c_\beta} \left[(\mbo^2 - \mbt^2) \sbt 
      + 2 m_b \mu t_\beta\right]\\
m_b^2\tilde{g}_{\tilde{b},11}^{H2} &= 
      \frac{1}{6} m_Z^2 s_\beta \left[3 + \cbt (1+2\ctw)\right]
      - \frac{\sbt}{c_\beta} m_b \mu \\
m_b\tilde{g}_{\tilde{b},11}^{H3} &= -\frac{1}{\sqrt{2}}\lambda v\sbt t_\beta\\
m_b^2\tilde{g}_{\tilde{b},12}^{H1} &= m_b^2\tilde{g}_{\tilde{b},21}^{H1} =
      \frac{1}{6}m_Z^2 \sbt c_\beta (1 + 2 \ctw) 
      + \frac{\cbt}{2c_\beta} \left[(\mbo^2 - \mbt^2) \sbt 
      + 2 m_b \mu t_\beta\right]\\
m_b^2\tilde{g}_{\tilde{b},12}^{H2} &= m_b^2\tilde{g}_{\tilde{b},21}^{H2} = 
      -\frac{1}{6}m_Z^2 \sbt s_{\beta} (1 + 2 \ctw) - \frac{\cbt}{c_\beta} m_b \mu \\
m_b\tilde{g}_{\tilde{b},12}^{H3} &= m_b\tilde{g}_{\tilde{b},21}^{H3} =
-\frac{1}{\sqrt{2}}\lambda v \cbt t_\beta\\
m_b^2\tilde{g}_{\tilde{b},22}^{H1} &= \frac{2 m_b^2}{c_\beta}
      - \frac{1}{6}m_Z^2 c_\beta\left[3 - \cbt (1+2\ctw)\right]
      - \frac{\sbt}{2c_\beta} \left[(\mbo^2 - \mbt^2) \sbt
      + 2 m_b\mu t_\beta\right]\\
m_b^2\tilde{g}_{\tilde{b},22}^{H2} &=  \frac{1}{6}m_Z^2 s_\beta \left[3 - \cbt (1+2\ctw)\right]
      + \frac{\sbt}{c_\beta} m_b\mu \\
m_b\tilde{g}_{\tilde{b},22}^{H3} &= \frac{1}{\sqrt{2}}\lambda v \sbt t_\beta
\end{align}}

{\allowdisplaybreaks
\begin{align}
m_t^2\tilde{g}_{\tilde{t},11}^{H1} &= \frac{1}{6}m_Z^2 c_\beta \left[3 + \ctt (-1 + 4 \ctw)\right]
      - \frac{\stt}{s_\beta} m_t\mu\\
m_t^2\tilde{g}_{\tilde{t},11}^{H2} &= \frac{2 m_t^2}{s_\beta} - \frac{1}{6}m_Z^2s_\beta\left[3 + \ctt (-1 + 4 \ctw)\right]
      + \frac{\stt}{2s_\beta} \left[(\mto^2 - \mtt^2)\stt + 2m_t\mu\frac{1}{t_\beta}\right]\\
m_t\tilde{g}_{\tilde{t},11}^{H3} &= -\frac{1}{\sqrt{2}t_\beta} \lambda v \stt\\
m_t^2\tilde{g}_{\tilde{t},12}^{H1} &= m_t^2\tilde{g}_{\tilde{t},21}^{H1} = 
      - \frac{1}{6} m_Z^2 \stt c_\beta (-1 + 4 \ctw)
      - \frac{\ctt}{s_\beta} m_t\mu \\
m_t^2\tilde{g}_{\tilde{t},12}^{H2} &= m_t^2\tilde{g}_{\tilde{t},21}^{H2} = 
      \frac{1}{6} m_t^2 \stt s_\beta (-1+4\ctw)  + \frac{\ctt}{2s_\beta}\left[(\mto^2 - \mtt^2) \stt + 2 m_t\mu\frac{1}{t_\beta}\right]\\
m_t\tilde{g}_{\tilde{t},12}^{H3} &= m_t\tilde{g}_{\tilde{t},21}^{H3} = -\frac{1}{\sqrt{2}t_\beta}\lambda v \ctt\\
m_t^2\tilde{g}_{\tilde{t},22}^{H1} &= \frac{1}{6}m_Z^2c_\beta\left[3 - \ctt (-1 + 4 \ctw)\right] +\frac{\stt}{s_\beta} m_t\mu \\
m_t^2\tilde{g}_{\tilde{t},22}^{H2} &= 
\frac{2 m_t^2}{s_\beta} - \frac{1}{6} m_Z^2 s_\beta  \left[3 - \ctt (-1 + 4 \ctw)\right]
-\frac{\stt}{2 s_\beta} \left[(\mto^2 - \mtt^2)\stt + 2 m_t\mu \frac{1}{t_\beta}\right] \\
m_t\tilde{g}_{\tilde{t},22}^{H3} &= \frac{1}{\sqrt{2}t_\beta}\lambda v\stt
\end{align}}

In the \cp{}-odd sector contributions with identical squark indices $\tilde{g}_{\tilde{q},ii}^{Ak}$ do not exist. 
The remaining ones are given by:
\begin{align}
m_b^2\tilde{g}_{\tilde{b},12}^{A1} &= -m_b^2\tilde{g}_{\tilde{b},21}^{A1} = \frac{1}{2c_\beta} \left[(\mbo^2 - \mbt^2) \sbt
       + 2 m_b \mu t_\beta\right]\\
m_b\tilde{g}_{\tilde{b},12}^{A2} &= -m_b\tilde{g}_{\tilde{b},21}^{A2}  = \frac{\mu}{c_\beta}\\
m_b\tilde{g}_{\tilde{b},12}^{A3} &= -m_b\tilde{g}_{\tilde{b},21}^{A3} = \frac{1}{\sqrt{2}} \lambda v t_\beta
\end{align}
\begin{align}
m_t\tilde{g}_{\tilde{t},12}^{A1} &= -m_t\tilde{g}_{\tilde{t},21}^{A1} = \frac{\mu}{s_\beta}\\
m_t^2\tilde{g}_{\tilde{t},12}^{A2} &= -m_t^2\tilde{g}_{\tilde{t},21}^{A2} = \frac{1}{2s_\beta }\left[(\mto^2 - \mtt^2) \stt + 2 m_t \mu\frac{1}{t_\beta}\right]\\
m_t\tilde{g}_{\tilde{t},12}^{A3} &= -m_t\tilde{g}_{\tilde{t},21}^{A3} =  \frac{1}{\sqrt{2}t_\beta}\lambda v
\end{align}

All occurrences of the soft-breaking parameters $A_t$ and $A_b$ were replaced by
their relation to the squark mixing angles $\theta_b$ and $\theta_t$.
Trigonometric functions are abbreviated through $s_x=\sin x, c_x=\cos_x$ and $t_x=\tan x$.


\end{document}